\documentclass{aa}  

\usepackage{graphicx}
\usepackage{multirow}
\usepackage[colorlinks=true,linkcolor=black,citecolor=blue,urlcolor=magenta]{hyperref}
\usepackage{txfonts}
\usepackage{float}
\usepackage{amsmath}

\begin{document}

   \title{Young system development in a cometary globule}

   \subtitle{An investigation into the eccentric disk around AT Pyx$^\star$ in terms of planet formation and interaction with its surrounding environment}

   \author{D. McLachlan
          \inst{1,2}
          \and
          C. Ginski\inst{2}\and
          J. Byrne\inst{2}
          \and
          C. Lawlor\inst{2}\and
          J. Campbell-White\inst{3}\and
          R. Claes\inst{3}\and
          B. Ren\inst{4}\and
          A. Sierra\inst{5,6}
        }
   \institute{Charles University, Faculty of Mathematics and Physics, Astronomical Institute, V Holešovičkách 2, 180 00 Praha 8, Czechia
        \and
            School of Natural Sciences, Center for Astronomy, University of Galway, Galway, H91 CF50, Ireland
        \and
            European Southern Observatory, Karl-Schwarzschild-Strasse 2, 85748 Garching bei München, Germany 
        \and
            P.E.S., Observatoire de la C\^ote d'Azur, Nice 06304, France
        \and
            Universidad Nacional Autónoma de México, Instituto de Astronomía, A.P. 70-264, Ciudad de México 04510, México
        \and
            Mullard Space Science Laboratory, University College London, Holmbury St Mary, Dorking, Surrey RH5 6NT, UK
            }

   \date{Received November 5, 2025; accepted February 13, 2026}

  \abstract
   {To understand the formation of planetary systems, it is necessary to observe and study systems at different evolutionary stages and in different environments. This paper presents new data and analyses of the AT Pyx system, a disk-hosting young star located in a cometary globule in the Gum Nebula. This radiation-driven structure is an unusual environment for observations of planet formation, and differs greatly from the low-mass star-forming regions disks are most commonly observed in.}
  {Aided by a collection of visual and spectroscopic data available for this system, our aim is to infer the possibility of embedded planets existing within the disk and how the system's environment may affect its disk morphology.}
  {Using data from the VLT's instruments XSHOOTER, ESPRESSO, and  most prominently SPHERE, along with data from ALMA, we made a variety of measurements (geometric, photometric, and otherwise) to characterise the observed disk features and attributes such as spiral arms and eccentricity. Mapping of the velocity components was also undertaken using the ALMA gas line data to characterise disk orientation and determine the likelihood that the system is experiencing a late-stage infall event.}
  {The disk is measured to have a position angle of 28.06$\pm0.02$$^{\circ}$ and an inclination of 42.5$\pm0.5$$^\circ$. The disk is found to be eccentric with measured $e\approx0.626$ when deprojected. Under the assumption that the formation of a planet is wholly responsible for the primary and secondary spiral arms, we find the mass of such a planet can range between 0.004 and 3 Jupiter masses. Measurements of the velocities associated with nearby globule cloud material return reasonable velocities for a late-stage infall event. We estimate the far-ultraviolet (FUV) field strength at AT Pyx's location to be low in comparison to other surveyed disks. We also find that AT Pyx is possibly a binary system.}
  {AT Pyx is the first disk within a cometary globule to be spatially resolved, and is now the first such disk to be investigated to this extent. The work of this paper could potentially be a first step into the further study of disks in the moderate FUV environment of the Gum Nebula and its globules.}


   \maketitle

\section{Introduction}
    \label{sec:intro}
 AT Pyx (IRAS 08267-3336, WRAY 15-220) is a young ($5.1^{+1.5}_{-1.0}$ Myr, \citealt{Ginski_2022}) T Tauri star \citep{HerczegandHillenbrand2014} located within the cometary globule CG22 \citep{SahuandSahu1992} in the Gum Nebula, most recently analysed in \cite{Ginski_2022}, which presented scattered light VLT/SPHERE H- and K-band and VLT/NACO L-band observations of its circumstellar disk. In the light of AT Pyx's non-intuitive appearance in scattered light, \cite{Ginski_2022} found two possible scenarios for disk orientation: 1) disk position angle $\approx 0^\circ$, i.e. the disk major axis is congruent with the horizontal in the image plane, and 2) disk position angle $\approx90^\circ$, i.e. the disk major axis is perpendicular to the horizontal. In scenario 1 the fitting of an ellipse to both bounding ansae introduces an offset between ellipse centre and star position of 55 mas, which cannot be explained by projection effects and suggests that the disk is eccentric, while in scenario 2 the disk can be modelled as non-eccentric, but the lack of strong signal for the disk ansae introduces ambiguity in modelling the disk boundaries. This paper follows up on the conclusions drawn by \cite{Ginski_2022} with an in-depth analysis of new data from SPHERE and ALMA, supplemented by XSHOOTER and ESPRESSO observations. 
   
   Many protoplanetary disk-hosting systems observed in scattered light are located in low-mass star-forming regions (e.g. \citealt{Avenhaus_2018}, \citealt{Rich2022}, \citealt{Ginski2024Chamaeleon}, \citealt{Garufi2024}), and hence cometary globules are  unusual environments in which to study planet formation. The most comparable large-scale study of disk structures is \cite{valegard2024spherevieworionstarforming}, who provide a survey of scattered light disks in Orion, where disks have strong interactions with their environment (e.g. the formation of proplyds, \citealt{Odell1994}, \citealt{Aru2024}). The Orion star-forming region is a high far-ultraviolet (FUV) environment in which \cite{valegard2024spherevieworionstarforming} probed the relationship between FUV flux and disk extent; they find no clear trend, but they observed no disks in locations with FUV flux higher than $\sim300G_0$. In this paper, FUV flux is quantified by the Habing unit, $G_0$.\footnote{Average FUV flux in the ISM $G_0\approx1.3\times10^{-3}$ (\citealt{Habing1968}, \citealt{Winter2018})}

   The Gum Nebula \citep{Gum1952} is identified as an intermediate structure between a classical H{\sc{ii}} region and a typical supernova remnant by \cite{ChanotandSivan1983}, and plays host to at least 32 cometary globules (\citealt{Kim_2005}, \citealt{Choudhury2009GumNebulaGlobules}). Cometary globules as defined in \cite{HawardenandBrand1976} derive their name from their distinct comet-like shape, with compact, dusty, opaque heads and trailing tails. Such globules are classified as precursors \citep{Reipurth1983} to Bok globules \citep{BokandReilly1947}, i.e. isolated dense clouds with masses of the order of 10--100 M$_\odot$ and sizes of the order of 0.1--1 pc. Such globules may form one to a few isolated low-mass stars similar to the Sun \citep{elmegreen1998observationstheorydynamicaltriggers}. With the assumption that their structure is a result of radiation-driven implosion under FUV radiation \citep{Bertoldi1989B}, similarly to Orion we observe a direct interaction between the broader interstellar environment and the conditions of star and disk formation. The three ionising sources in the Gum Nebula, $\gamma^2$ Vel, $\zeta$ Pup, and the progenitor to the Vela supernova remnant Vela XYZ (\citealt{Kim_2005}, \citealt{Choudhury2009GumNebulaGlobules}), contribute to the moderate FUV environment sufficiently to photoevaporate the globules into their characteristic shapes. The FUV fluxes calculated at globule positions ($6.6^{+3.2}_{-2.7}$ $G_0$ at CG30 from \citealt{Yep2020}, between 0.93 and 28.74 $G_0$ at CG22; see Appendix \ref{sec:fuv}) are low in comparison to the high FUV fluxes calculated in the Orion Nebula Cluster ($\sim$30000$G_0$ \citealt{Winter2018}) or at the locations of some of the stars surveyed by \cite{valegard2024spherevieworionstarforming} with FUV flux higher than 300$G_0$. This makes it  difficult to determine the extent to which the Gum Nebula's moderate FUV environment shapes the evolution of circumstellar disks. 

   In this paper we present new analyses of the AT Pyx system based primarily on VLT/SPHERE, VLT/XSHOOTER, and ALMA observations of AT Pyx (supplemented by VLT/ESPRESSO observations described in Appendix \ref{sec:espresso})  to probe the planet-forming potential of the disk as well as its interactions with its surrounding globule environment. This paper is structured as follows. In section \ref{sec:observations} the observations of AT Pyx are described in detail. Section \ref{sec:methods} describes the methods used to probe the signatures of embedded planets within the disk. Section \ref{sec:results} details the resulting characteristics of a hypothetical planet needed to drive the observed features. Section \ref{sec:discussion} elaborates on other non-planet-related processes speculated on within the disk. Section \ref{sec:conclusions} provides our conclusions from this paper.

\section{Observations}
\label{sec:observations}

\subsection{XSHOOTER/spectroscopic observations}
\label{sec:xshooter}

 XSHOOTER spectra (ESO programme ID: 112.25BZ.001) were interpreted using the methodologies described in \cite{Claes2024} to determine the stellar parameters in Table \ref{tab:xshooterproperties}. We used data from two observing epochs, 2023 September 25 and 2023 November 06, where there are slight deviations in the calculated values. The stellar mass and accretion rate were obtained using the stellar evolution models of \cite{baraffe2015}, which provide a mass range of 1.216--1.283 M$_\odot$. We used the typical uncertainties associated with this method, as described in \cite{manara2023}.

Additional spectroscopic data from ESPRESSO are shown in Appendix \ref{sec:espresso}, but are not included in the main text. The information obtained from the ESPRESSO spectra is used in investigating forbidden line emission, which does not inform the primary analytical conclusions of this work.

{\renewcommand{\arraystretch}{1.2}
 \begin{table} 
    \setlength{\tabcolsep}{2.5pt}
    \centering
    \caption{Stellar properties obtained from XSHOOTER spectra for AT Pyx. }
    \begin{tabular}{c|cc|c|c}
    \hline \hline 
    Parameter & 2023/09/25 & 2023/11/06 & Error & Unit\\
    \hline

    Distance & 370 & 370 & 5 & pc\\

    SpT & K2.5 & K2.5 & 0.5 & -\\
    
    T\textsubscript{eff} & 4625 & 4625 & 100& K\\
    
    log(L$_\star$)  & 0.04 & -0.035 & 0.04 & L$_\odot$\\
    A$_V$  &1.5 &1.7 & 0.3 & mag\\

    M\textsubscript{$\star$}  & 1.278 & 1.221&0.05&M$_\odot$\\
    
    $\log(\dot M_{\text{Acc}})$  & -8.46 & -8.51&0.35&M$_\odot$/yr\\

    \hline \hline
    \end{tabular}
    \tablefoot{GAIA DR3 \citep{DISTANCE_gaia_dr3} distance included here. For all XSHOOTER parameters, uncertainties are given as 3$\sigma$ limits.}
    \newline
    
    \label{tab:xshooterproperties}
\end{table}
}
\begin{figure}  
 \centering
\includegraphics[width=0.95\hsize]{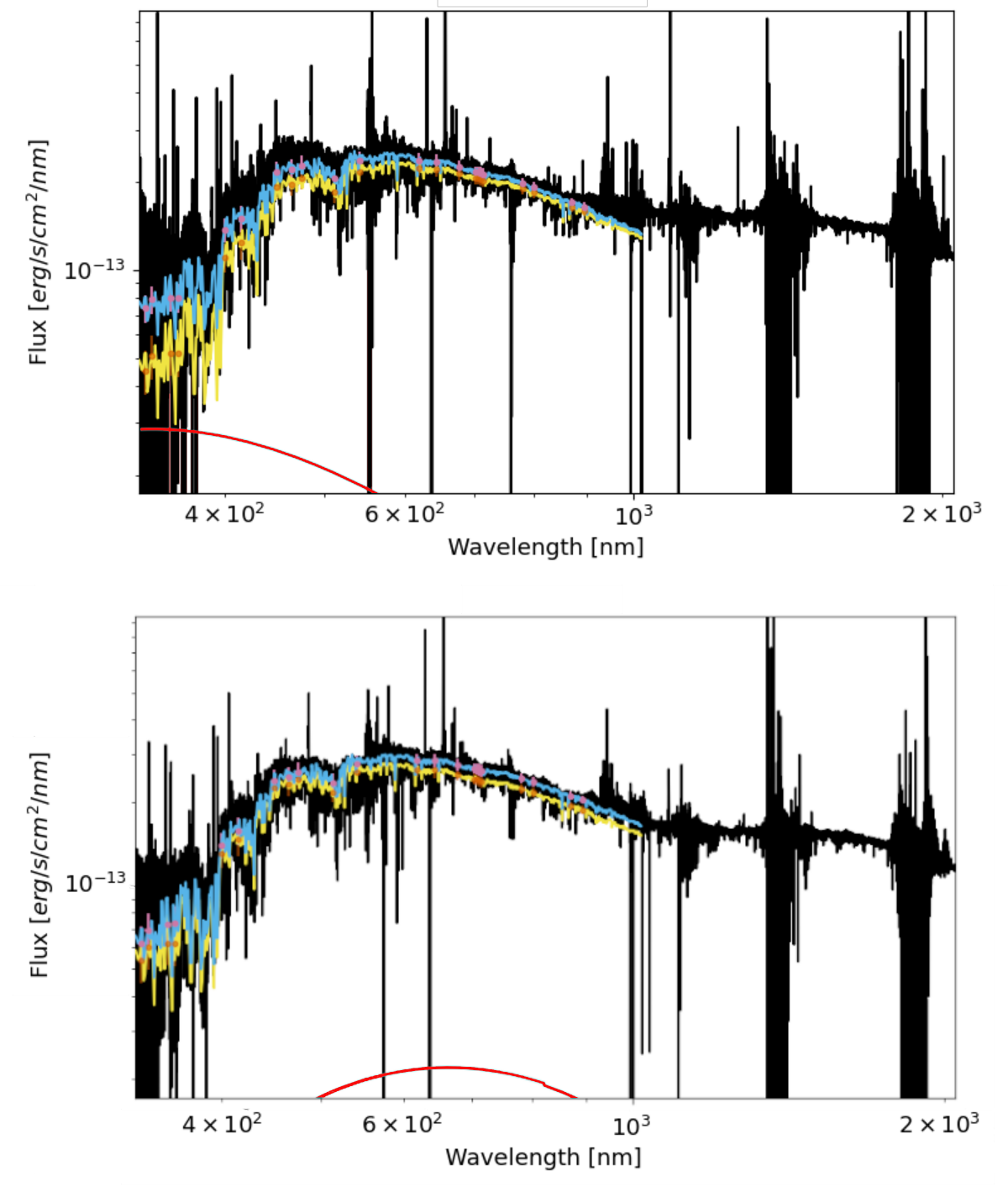}
\caption{Characteristic stellar spectra for AT Pyx. Black indicates the observed spectrum, yellow the best fit photospheric template, red the slab model, light blue the best fit. \textit{Top:} 2023 September 25 spectrum. \textit{Bottom:} 2023 November 06 spectrum.}
\label{fig:xshooterspectrum}
\end{figure}

\subsection{ALMA}
\label{sec:alma}
{\renewcommand{\arraystretch}{1.2}
\begin{table*}

\centering
    \caption{ALMA image and cube parameters}
    \label{tab:ALMAImaging}
    \begin{tabular}{c|ccccc}
    \hline \hline 
    \multirow{2}{*}{Data}  & \multirow{2}{*}{Robust} & Beam Size & rms & Vel resolution \\
    & & (arcsec) & (mJy beam$^{-1}$) &(m s$^{-1}$) & \\
    
    \hline
    Continuum & 0.5 &  0.539$\times$0.397 & 0.3 & - \\
    $^{12}$CO (J=2-1) & 0.5 & 0.539$\times$0.397 & 4.2 & 650 \\
    
    \hline \hline
    \end{tabular}
\end{table*}
}

AT Pyx's disk was imaged with ALMA at Band 6 (230GHz) in the molecular gas line \textsuperscript{12}CO (J=2-1) and in 1.3 mm dust continuum emission. The  beam size for both datasets was 539x397 mas$^2$ after reprocessing (ALMA Project ID: 2021.1.01705.S, observing epoch 2022 August 31; see table \ref{tab:ALMAImaging} for image and cube parameters). 

\subsubsection{Millimetre dust continuum emission}
\label{sec:almacontinuum}
\begin{figure}  
   \centering
   \includegraphics[width=\hsize]{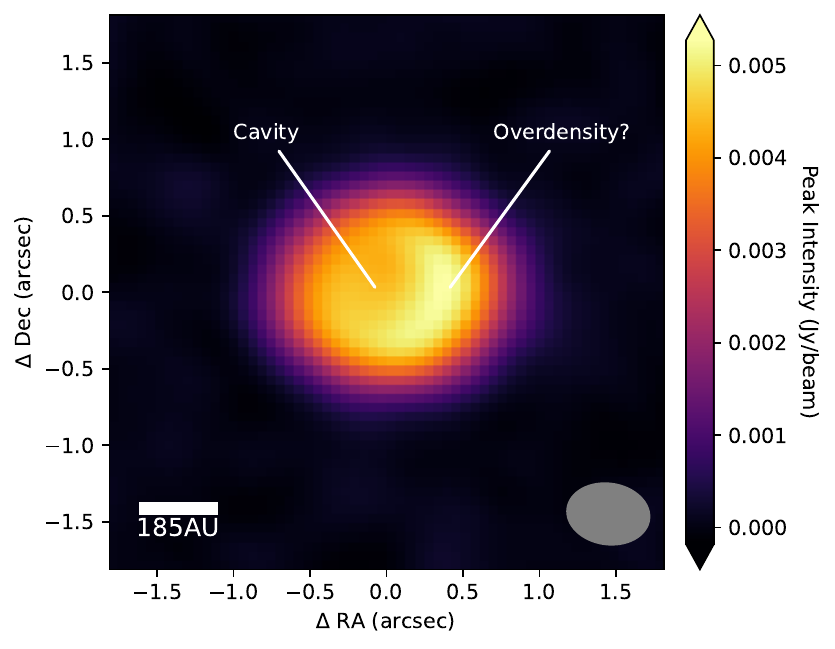}
      \caption{ALMA 1.3 mm dust continuum image. The grey ellipse indicates beam size and angle.
              }
         \label{fig:continuum}
   \end{figure}
The resolved dust continuum image appears to show a cavity in the central region (see Fig. \ref{fig:continuum}), and when we overlay the SPHERE scattered light H-band image on the continuum image it is notable that the bright arcs seen in the east and west of the disk appear to trace the cavity region (see Appendix \ref{sec:cavity}). Continuum emission extends out to $\sim$1'' (370 au), roughly half as extended as the peak intensity \textsuperscript{12}CO emission ($\sim$2'', 740 au). 

From the millimetre flux obtained from aperture photometry of the dust continuum image, the disk dust mass can be calculated using the following relation \citep{Hildebrand_1983}: \begin{equation}M_d=\frac{F_{\nu}D^2}{B_{\nu}(T)\kappa_{\nu}}.\end{equation} Here $F_\nu$ is the integrated dust continuum flux, D the distance to the system, and $B_\nu$ the Planck function  expressible in the Rayleigh-Jeans form $B_{\nu} \approx2\nu^2kT/c^2$ in millimetre wavelengths \citep{Williams_2011}. The disk temperature can be estimated as $T=43(L/L_{\odot})^{0.25}\approx43$ K \citep{Kristensen_2012} as AT Pyx was found to have a luminosity of $\sim$1 L$_{\odot}$ from the XSHOOTER observations. The dust opacity, \begin{equation}
    \kappa_{\nu}=2.3\left(\frac{\nu}{230\text{GHz}}\right)^{\beta}\text{cm}^2\text{g}^{-1},\end{equation} varies with grain size distribution as represented by the dust spectral index $\beta \approx 0 - 2$ (\citealt{Hartmann_2008}, \citealt{Andrews_2013}; see also \citealt{Sheehan_2022}). In this case $\beta=1$ is used giving a dust opacity of 2.3 cm$^2$g$^{-1}$. A system distance of 370 pc is obtained from Gaia DR3 \citep{DISTANCE_gaia_dr3}.

The integrated flux density is obtained using aperture photometry of the dust continuum image to sum up all the pixel flux density values within the region of continuum emission. The summed values are then converted from Jy/beam to Jy by dividing this figure by beam solid angle,  \begin{equation}
\Omega_b=\frac{\pi}{4\ln2}\times b_{maj}\times b_{min},\end{equation} where the beam major and minor axes are in units of pixels. The final integrated flux density is measured as $F_{\nu}=24.88\pm 2.4$ mJy, and thus the disk dust mass is calculated to be $97\pm9$ M$_\oplus$. Assuming a 1:100 relationship between dust and gas mass in circumstellar disks  yields a total disk mass of $9700\pm900$ M$_\oplus$ or $30.5\pm2.8$ M\textsubscript{Jup}.

\subsubsection{\textsuperscript{12}CO (J=2-1) emission}
\label{sec:almagas}
\begin{figure*}
   \resizebox{\hsize}{!}
            {\includegraphics[width=0.32\linewidth]{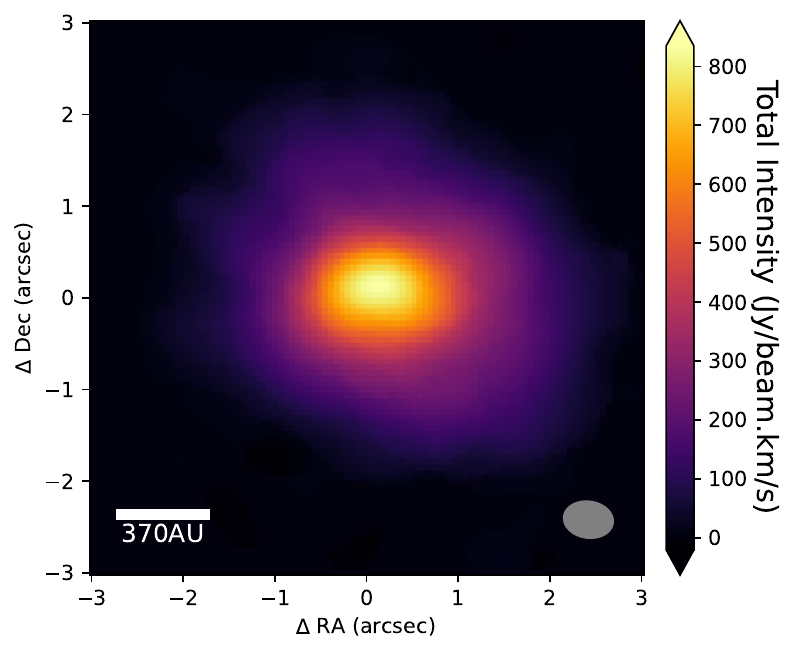}
            \includegraphics[width=0.32\linewidth]{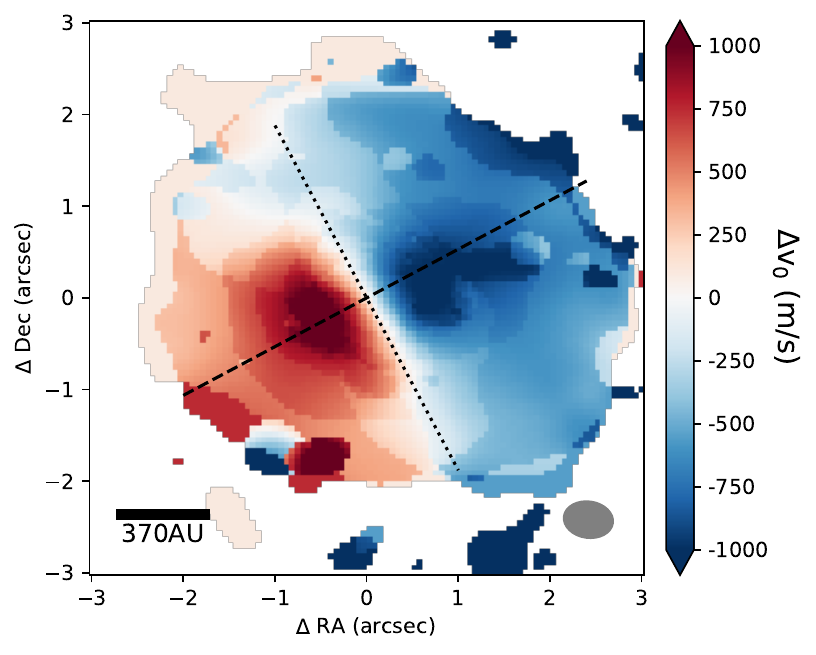}
            \includegraphics[width=0.32\linewidth]{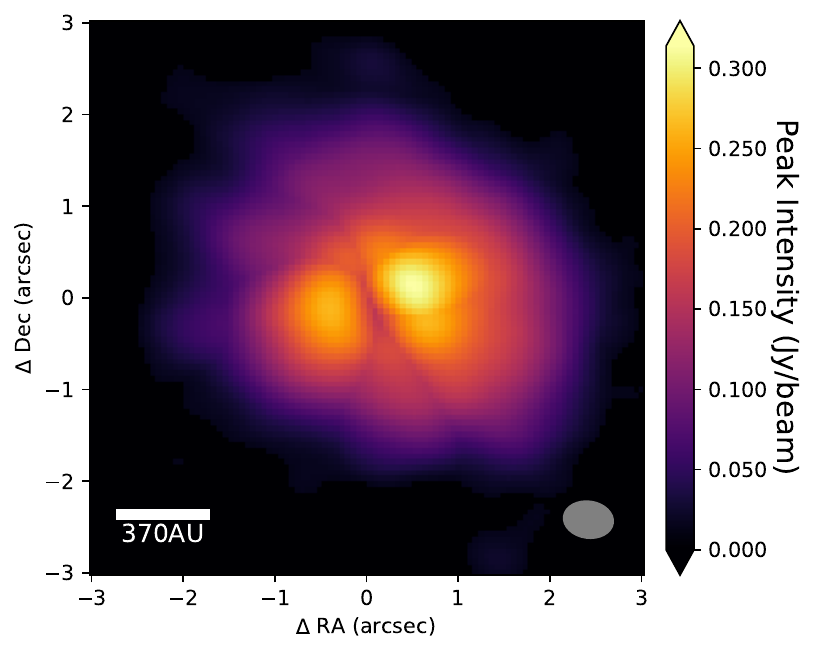}
            }
      \caption{ALMA images of the AT Pyx system in \textsuperscript{12}CO (J=2-1) line emission. \textit{Left:} Total intensity image (Moment 0). \textit{Middle:} Intensity-weighted velocity (Moment 1). The major (dashed) and minor (dotted) axes are overplotted in black. \textit{Right:} Peak intensity (moment 8). 
              }
         \label{fig:almamoments}
   \end{figure*}

From the moment 1 maps of the gas line data it can be seen that the disk major axis (perpendicular to the dividing line between the redshifted and blueshifted sections) lies angled slightly against the image x-axis (west--east on-sky; see figure \ref{fig:almamoments}). This validates scenario 1 for the disk's on-sky orientation from \cite{Ginski_2022}. 

The brightness distribution in \textsuperscript{12}CO emission as seen in the total and peak intensity maps (see Fig. \ref{fig:almamoments}) is broadly asymmetric, much like the brightness distribution in continuum emission. The gas line data is discussed further in the analyses in Sect. \ref{sec:eddy}.

\subsection{SPHERE/scattered light}
\label{sec:sphere}

\begin{figure*}
   \resizebox{\hsize}{!}
            {\includegraphics[trim={0 0 0 1.4cm},clip,width=0.4\linewidth]{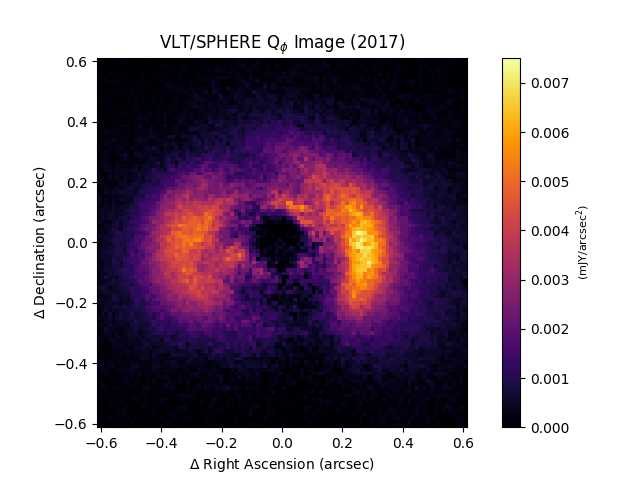}
            \includegraphics[trim={0 0 0 1.4cm},clip,width=0.4\linewidth]{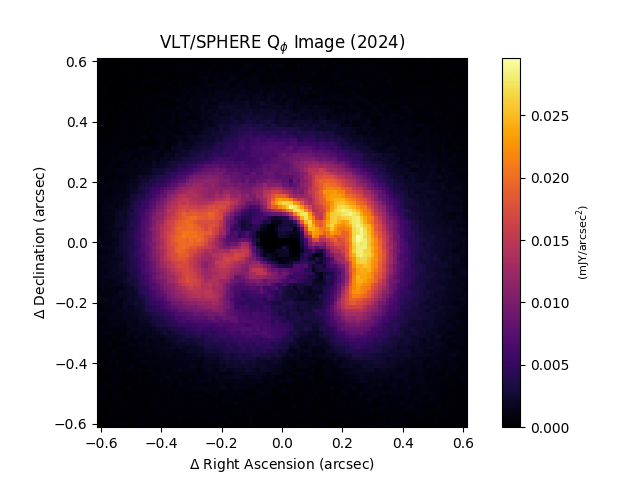}}
      \caption{SPHERE H-band Q$_\phi$ images from 2017 (left) and 2024 (right). All disk features are identical between epochs;  the only difference is the better signal-to-noise ratio in the 2024 observation.
              }
         \label{fig:sphereimages}
   \end{figure*}

\begin{figure}  
   \resizebox{\hsize}{!}
            {\includegraphics[width=0.99\linewidth]{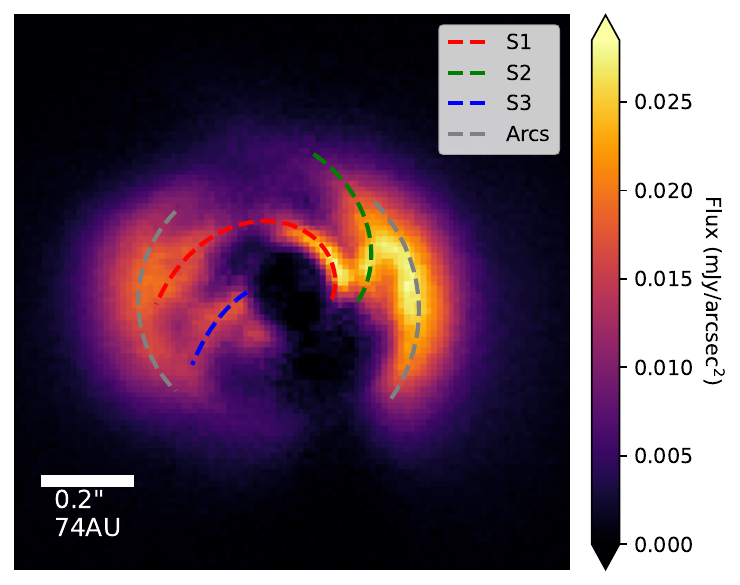}}
      \caption{AT Pyx disk features in H-band scattered light SPHERE Q$_\phi$ image. The spiral features are numbered in order of decreasing apparent length. North is up and east is to the left. 
              }
         \label{fig:bigsphereimage}
   \end{figure}

 SPHERE/IRDIS observations were taken on January 17, 2024 (ESO programme ID: 112.25B7.003). These served to improve on a previous set of observations taken on May 15, 2017 (ESO programme ID: 099.C-0147(B)), with a lower signal-to-noise ratio on which the findings of \cite{Ginski_2022} were based (see figure \ref{fig:sphereimages}). 

The 2024 observations were taken with seven polarimetric cycles,  each containing four half-wave cycles, while the 2017 observations were taken with three polarimetric cycles; an integration time of 96 seconds/image was used for both sets of observations. In both cases a coronagraph with a 92.5 mas inner working angle was used to block stellar light. Average seeing for the 2024 observations was 0.573, superior to the 2017 observations' average seeing of 0.676. This provides a better signal-to-noise ratio for the newer observations; as AT Pyx has a Gaia magnitude of 13.319$\pm0.02$ \citep{DISTANCE_gaia_dr3}, which is at the lower end of brightness that can be observed successfully with SPHERE,  observations of AT Pyx are particularly sensitive to the quality of atmospheric seeing. The residuals seen when subtracting the images from each other in Appendix \ref{sec:spherespiralmovement} are minimal and appear to only be the result of better adaptive optics and signal-to-noise in the newest observing epoch rather than any changes in the system detectable within a SPHERE beam. For all images, data reductions were performed with  IRDIS Data reduction for Accurate Polarimetry (IRDAP; \citealt{irdapvanholstein_2020}) using the default settings. Presented in figure \ref{fig:sphereimages} are the final Q$_{\phi}$ images.

In \cite{Ginski_2022} three spiral features were highlighted from the 2017 dataset; they can be identified consistently across both observing epochs. As all distinct disk features are identical in the two observational epochs, we present the 2024 Q$_\phi$ image with all distinct features highlighted in Fig. \ref{fig:bigsphereimage}. In Appendix \ref{sec:cavity}, we also show how the arcs seen in scattered light trace the cavity edges in continuum emission.

\section{Methods}
\label{sec:methods}
\subsection{Gas velocity analysis}
\label{sec:eddy}
In order to proceed with a morphological analysis of AT Pyx's disk we acquired information on the disk's 3D structure from the ALMA gas line data, most prominently values for disk position angle and inclination. To this end we generated rotation maps using the package Eddy \citep{eddy} where Keplerian rotation patterns are fitted to peak velocity (gv0) maps of AT Pyx created with the BetterMoments package \citep{bettermoments}.

Our goal was thus to find the solution that provided the most accurate values for the known parameters and so the concurrent values for the unknown parameters must be accurate and could be used in further analysis of the higher-resolution scattered light image. Parameters for Eddy fitting are listed in Appendix \ref{sec:eddycorner}. The disk position angle was measured using Eddy's fitting functionalities, as can be seen in Appendix \ref{sec:eddycorner} (see Fig. \ref{fig:eddycorners} for the value; see \ref{fig:positionangle} for a visual representation of the measured position angle). This value began as a rough guess of $\sim$ 22$^\circ$, evolving into the final position angle value of 28.06$\pm0.02$$^{\circ}$ through the process of Eddy fitting. A systemic velocity value of $-12.1\pm0.4$ km/s was acquired from fitting a Gaussian to the velocity channels of the ALMA datacube. We acquired accurate stellar mass values within the range of 1.22--1.28 M$_{\odot}$ from the XSHOOTER data.

To obtain a value for disk inclination, a number of runs were performed while varying input inclination. We used an initial range of inclination values between 35$^{\circ}$ and 45$^{\circ}$ (based on the values from \citealt{Ginski_2022}),  and iteratively narrowed the range depending on what input value produced the most accurate stellar mass.

\begin{figure*}
    \sidecaption
    \includegraphics[width=6cm]{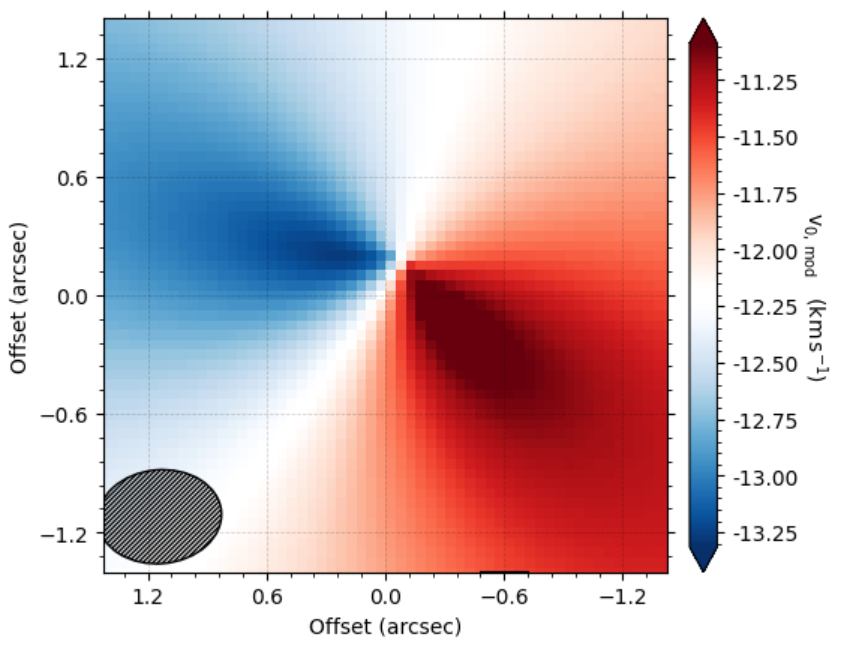}
    \includegraphics[width=6cm]{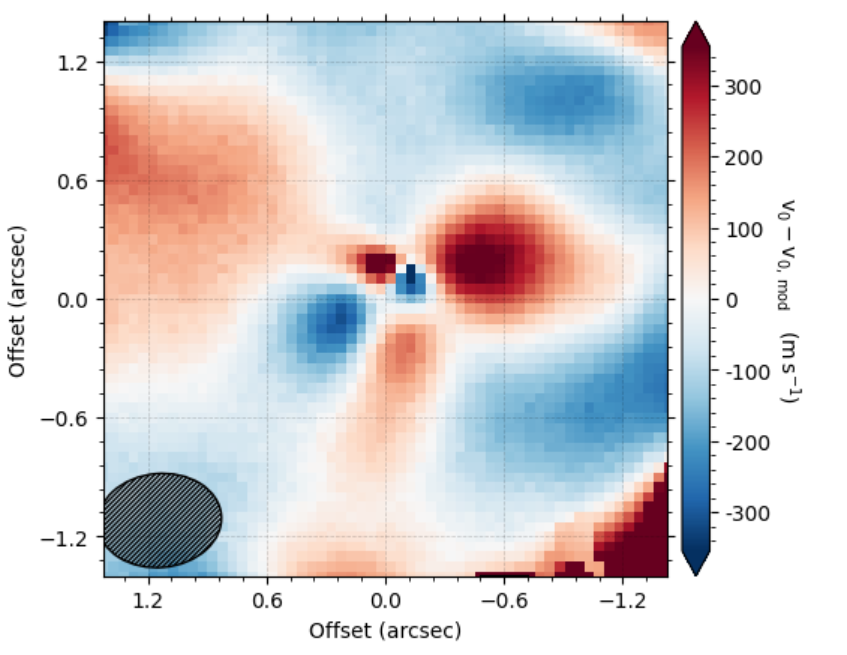}
    \caption{\textit{Left:} Rotation map showing a model Keplerian rotation pattern for AT Pyx's gas content. \textit{Right:} Residual, i.e. the rotation map subtracted from the peak velocity image,   showing deviations in the rotation patterns from a Keplerian velocity field. We note that the high residuals (dark red) could indicate some level of pollution from the \textsuperscript{12}CO signal in the surrounding globule gas. In these images north is up and east is to the right.}
\label{fig:eddyplots}
\end{figure*}

Figure \ref{fig:eddyplots} shows the final model and residual map, and the final converged model parameters can be seen in Appendix \ref{sec:eddycorner}, Fig. \ref{fig:eddycorners}. An inclination of $42.5$$^{\circ}$ was found to provide the closest value for stellar mass to the known range at 1.2295$\pm$0.0019 M$_{\odot}$. The model mass deviated significantly from the XSHOOTER mass for inclinations with a 0.25$^{\circ}$ difference from 42.5$^{\circ}$, and hence this is used as the error on this value. Referring to Fig. \ref{fig:bigsphereimage} this means that the disk major axis is at 28.06$\pm0.02$$^{\circ}$ to the x-axis and the disk is inclined at 42.5$\pm0.25$$^{\circ}$ to the observer's line of sight;  the north edge is most likely the near side as more flux is received from this side in polarimetric scattered light observations.

A common use of Eddy is to find agreement between deviations from Keplerian rotation patterns and observed features in scattered light (e.g. \citealt{Teague_2019} in the case of TW Hya). However visible tracers cannot be accurately represented due to the low spatial and spectral resolution of the ALMA data (see Table \ref{tab:ALMAImaging}). The gas emission is covered in Eddy runs by a 50x50 pixel bounding box in order to mask out noisy regions, which is only $\sim8\times6$ beams. The presence of additional non-disk signal due to contamination from the surrounding globule envelope could also be responsible for the large residuals (see figure \ref{fig:eddyplots}); the velocity channels approaching systemic velocity play host to cloud signal, and there are not enough available channels to strip contaminated channels out without losing valuable disk signal. 

Using the values for position angle and inclination found in this section, a deprojected image of the disk was prepared based on the disk height profile equation \begin{equation}
    \label{eqn:scaleheight}
h(r) =h_{0}\left(\frac{R}{r_{0}}\right)^{\alpha},
\end{equation} with sample scale height $h_0$ at 100 au of 16.17 au and flaring power $\alpha$ of 1.219 obtained from \cite{Avenhaus_2018}. The deprojection code was validated on the more `well-behaved' disk in the WISPIT 2 system (see \citealt{vancapelleveen2025}), and this deprojected image is used in sections \ref{sec:armstructure} and \ref{sec:eccentricity} (see figures \ref{fig:spiraldeproj} and \ref{fig:ellipsefit}).

\subsection{Spiral modelling}
\label{sec:spiralarms}
The presence of spiral features may not necessarily always indicate the presence of planetary perturbers. Large spiral arms can be generated in disks by self-gravitating forces from global gravitational instability (\citealt{Lin1964}, \citealt{Dong_2016SpiralInstability}, \citealt{Baehr_2021}). However, in Appendix \ref{sec:gravitationalinstability} it is found unlikely that the disk is gravitationally unstable. An attempt was made to test the motion of spiral arms between SPHERE observing epochs, which is detailed in Appendix \ref{sec:spherespiralmovement} and was not successful. In this section it is assumed that the spirals are driven by gravitational influence from an embedded planet in order to fully investigate the planet-forming scenario. In addition, with new higher signal-to-noise SPHERE observations a more comprehensive analysis on spiral structure could be undertaken than  was possible in \cite{Ginski_2022} with the 2017 SPHERE dataset mentioned in Sect. \ref{sec:sphere}. In this section the photometric contrast ratios of the spiral arms, radially dependent pitch angles of spiral arms, and azimuthal arm-to-arm separations are used to constrain the mass of a possible embedded planetary-mass companion.

\subsubsection{Spiral arm contrast}
\label{sec:armcontrast}
The analysis in this section draws heavily from the work of \cite{Dong_2017} (hereafter DF17) in which hydrodynamical and radiative transfer simulations of disks with a range of masses of planetary perturbers at a fixed radial distance demonstrated a relationship between contrast of spiral arms and planet perturber mass. In the simulation results from DF17 each mass drives a signature peak contrast at either side of the planet's orbital radius. Contrast in DF17 is defined as \begin{equation}\delta(r) = \frac{\text{Max Surface Brightness}(r)}{\text{Azimuthally Averaged Surface Brightness}(r)}\end{equation} and in the case of AT Pyx we identify three features (S1, S2, S3) as spiral arm candidates (see figure \ref{fig:bigsphereimage}), so for each spiral the contrast equation can be modelled as \begin{equation}\delta(r) = \frac{\text{Spiral Brightness}(r)}{\text{Azimuthally Averaged Surface Brightness}(r)}.\end{equation}

Hence, each spiral's contrast as a function of radial distance can be modelled and compared to provide some provisional values for planet perturber mass. We fit each spiral with a Galilean logarithmic spiral of the form $r=R_0\exp(b\theta)$ to provide pixel co-ordinates from which to extract brightness values. Using the photutils library \citep{larry_bradley_photutils_2025_14889440}, an aperture simulating a $49\times49$ mas SPHERE beam extracts flux values along the co-ordinates of the spiral, where  each brightness has an associated deprojected radial distance from the central star's position. At each radial distance an elliptical annulus 49 mas thick with semi-major axis $a=R$, semi-minor axis $b=R\sin(\pi/2 -\text{inc})$, and the major axis lying along the disk major axis finds the azimuthally averaged surface brightness at each spiral co-ordinate. Dividing the spiral brightnesses by the azimuthally averaged brightnesses yields the radially dependent contrast for each spiral arm.

\subsubsection{Spiral arm structure and distribution}
\label{sec:armstructure}
In the domain of linear wave theory \citep{GoldreichandTremaineLinearWaveTheory1979}, planet-driven spiral arms in a circumstellar disk are understood to be formed by constructive interference between density waves with different azimuthal wavenumbers $m$ excited by the perturbing planet at Lindblad resonances \citep{Ogilvie_2002}. From this model for spiral density waves, the following relation can be used to predict the shape of the gravitational wake outside planet position $(R_p,\phi_p)$ in polar co-ordinates (\citealt{Rafikov2002}, \citealt{Muto2012}),

\begin{equation}
    \phi (R) = \phi_p - \frac{\text{sgn}(R-R_p)}{(h/r)_p}\nonumber
\end{equation}
\begin{equation}
    \label{eqn:spiralphase}
    \times \left[\left(\frac{R}{R_p}\right)^{1-\eta}\left(\frac{1}{1-\eta}-\frac{1}{1-\zeta+\eta}\left[\frac{R}{R_p}\right]^{-\zeta}\right)-\left(\frac{1}{1+\eta}-\frac{1}{1-\zeta+\eta}\right)\right],
\end{equation}

with orbital frequency $\Omega(R)\propto R^{-\zeta}$ and sound speed $c_s(R) \propto R^{-\eta}$.
This linear wave theory-based model has been shown to break down for all but very low-mass ($\sim$0.01 M\textsubscript{Jup}) wave-generating planets, but on disk scattering surfaces the outer spiral arms still broadly follow linear wave theory, while the inner arms are more open than predicted by \cite{Zhu_2015}. As AT Pyx lies further away than other similar disks, much more of the inner disk material is obscured by the coronagraph, making it difficult to probe the inner spiral arm structure completely, but it is possible to inspect how closely the visible and/or outer portions of the arms follow linear wave theory in order to infer the location of a planetary perturber. 

Furthermore, with regard to spiral behaviour according to linear wave theory, \cite{Rafikov2002} gives the following relationship between spiral pitch angle, scale height at planet radial distance, and sound speed:
\begin{equation}
\label{eqn:pitchangle}    
\tan\theta = r\frac{\Omega(r)-\Omega_p}{c_0(r)}=\frac{r_p}{h_p}\left(\frac{r}{r_p}\right)^{\nu+1}\left[\left(\frac{r_p}{r}\right)^{3/2}-1\right].
\end{equation}
From this it can be seen that the spiral opening angles increase with aspect ratio and are not governed solely by mass and position of a planetary perturber.

With regard to planet--disk interaction generating spirals, $M_p$ can be described in terms of disk thermal mass 
\begin{equation}
\label{eqn:thermalmass}
M_{th}=\frac{c_s^3}{G\Omega_p}=M_*(h/r)_p^3.
\end{equation} 
When planet mass approaches or exceeds $M_{th}$ such a planet can excite non-linear density waves that produce multiple spiral arms (\citealt{LinPapaloizou1993}, \citealt{GoodmanRafikov2001}).

The azimuthal angular separation of spiral arms can also be used to infer perturbing planet properties; \cite{Fung__2015spiralangleseparation} found a scaling relationship between azimuthal primary-to-secondary arm separation as 
\begin{equation}
    \label{eqn:fungdongseparation}
    \phi_{p-s}=102^{\circ}(q/0.001)^{0.2},
\end{equation}
where $q=M_p/M_*$, the planet/star mass ratio. This relation holds for high-mass planets located outside the observed spiral arm system, and thus can be used to describe the same hypothetical planet described in  section \ref{sec:armcontrast} above. This is further refined by \cite{Bae_2018b}, who define a wider range of fits in cases with varying $(h/r)_p$, and also provide limits on numbers of spiral arms generated although in the case of AT Pyx's disk, low spatial resolution and obfuscation of the inner disk material by the coronagraph significantly limits any conclusions that can be drawn about the number of spiral arms in the system. Hence this paper focuses only on the three spiral features observed here. A similar approach to this can be found in \cite{Reggiani_2018}, where the azimuthal separation of spiral arms is used to constrain the masses of hypothetical planetary companions in the system MWC 758 (see Fig. 7 therein).

\begin{figure}  
 \centering
\includegraphics[width=\hsize]{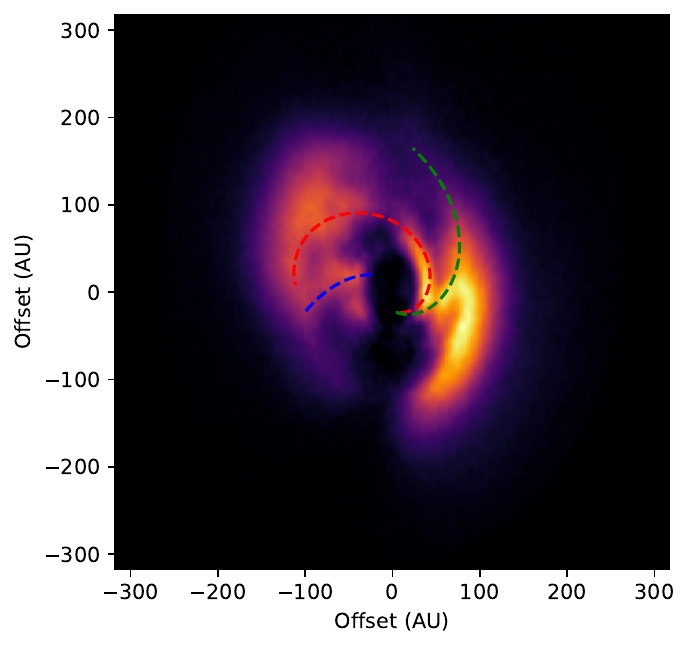}

\caption{Spirals modelled on deprojected disk image. Key: S1 - Red, S2 - Green, S3 - Blue. S2 and S3 can be modelled as standard Galilean spirals, whereas S1 deviates significantly from the default Galilean spiral shape.}
\label{fig:spiraldeproj}
\end{figure}

The deprojected spiral arms (see Fig. \ref{fig:spiraldeproj}) were modelled as Galilean spirals of the form $r = R_0\exp(b\theta)$ with $R_0$ being distance between (0,0), i.e. the star position, and the starting point of the modelled spiral in au. Starting points for each spiral may exist within the region masked by the coronagraph, but measurements are only performed on the visible regions, so functionally each spiral's starting point is at 32 au from the centre. With regard to S2 and S3 the spiral formulae need no modification to fit when overplotting on the deprojected image and are as follows: $$\text{S2: }r = 23\exp(0.7\theta),$$ $$\text{S3: }r = -31\exp(1.3\theta).$$ S1 requires constraints to reshape the spiral to overplot effectively as follows: $$r = 25\exp(0.38\theta),$$ $$x = r\cos(\theta)+0.06\theta^4,$$ $$y = r\sin(\theta)+0.12\theta^4.$$ Additionally S3's status as a true spiral arm is ambiguous as it appears to follow a path almost directly radially outwards from the disk centre and arguably an overplotted straight line could fit to it when deprojected. Nonetheless, its pitch angle is measured here alongside its companions. The spiral pitch angle can be measured as a function of radius by finding the angle between the tangent at each point and the complement of the radial direction. This is the same property defined by Rafikov's equation (equation \ref{eqn:pitchangle}), and thus Rafikov's model can be fit to S1 (here assumed to be the primary arm) with the input parameters ($h_p,r_p,\nu$) varied until the model fits. 

To find the azimuthal separation with increasing radial distance, the azimuthal positions of S1 (primary arm) and S2 (secondary arm) are overplotted (see figure \ref{fig:spiralangles}). In \cite{Bae_2018b} other fits for angular separation to the planet/star mass ratio (modifications of equation \ref{eqn:fungdongseparation}) are provided for $r=0.2r_p,0.3r_p,0.4r_p,0.5r_p,0.6r_p$ and in this case it is possible to retrieve separations from 0.4$r_p$--0.6$r_p$. The fits used here are 
$$r=0.4r_p:\phi_{p-s}=109^{\circ}(q/0.001)^{0.22},$$
$$r=0.5r_p:\phi_{p-s}=104^{\circ}(q/0.001)^{0.19},$$
$$r=0.6r_p:\phi_{p-s}=104^{\circ}(q/0.001)^{0.14}.$$

\subsection{Eccentricity}
\label{sec:eccentricity}

In \cite{Ginski_2022} it was found that constraining the fit of an ellipse to encompass both the east and west ansae of the disk put the star at an offset of 55 mas from the ellipse centre, assuming a position angle of 0$^{\circ}$ between disk midplane and image x-axis (not too far removed from the position angle found in section \ref{sec:eddy}). Such an offset could indicate that the disk is intrinsically eccentric. 

Eccentric features in disks have been observed previously; in the cases of HL Tau \citep{ALMAPARTNERSHIP_Brogan_2015} and HD 163296 \citep{Isella2016} disks were seen to have eccentric gap edges traced possibly as a result of embedded planets. Additionally, \cite{kley2005} demonstrate that in two-dimensional gas-only hydrodynamical simulations of viscous disks ($\alpha = 4\times10^{-3}$)  with embedded high-mass planets, a maximum disk eccentricity of $\sim0.25$ peaking at 1.2$r_p$ was found for a 5 M$_{Jup}$ planet after 2500 orbits.

A more in-depth study by \cite{Zhang_2018} (hereafter Zhang18) provides a range of 2D  simulations of millimetre dust continuum and gas surface density images, which show a trend towards greater gap eccentricity with lower viscosity, lower aspect ratio, and higher planet mass. The gas surface density simulations can be assumed to apply to scattered light images as small dust grains are well coupled to the disk gas content \citep{Takeuchi_Artymowicz_2001}. In section \ref{sec:almacontinuum} (and shown further in Appendix \ref{sec:cavity}) we suggest the ansae or arcs seen in the SPHERE scattered light image of the disk may trace the edges of the dip in brightness in dust continuum emission, so these ansae could be the edges of a ring bordering a gap between the inner and outer disks, in which case the Zhang18 models could apply well in relating disk eccentricity to an embedded planet. However, it is difficult to establish full continuity between observing regimes at the resolutions imposed by the system distance.

\begin{figure}  
 \centering
\includegraphics[width=0.8\hsize]{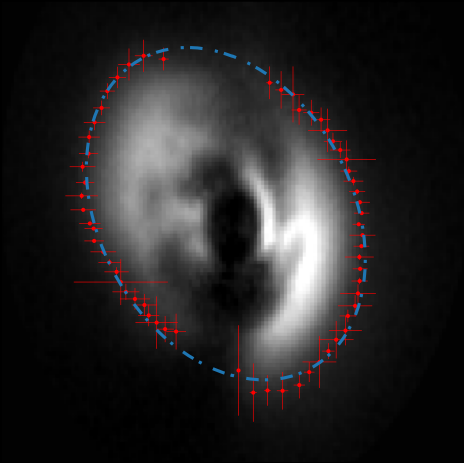}

\caption{\textit{Top:} Ellipse fitted to deprojected scattered light disk image. Red indicates the furthest-out edge points detected above a 5$\sigma$ threshold while also satisfying a three-point step function. For the sake of clarity, only every fifth point is displayed. Blue indicates a non-constrained least-squares ellipse fitted to the detected points. The eccentricity in this case was found to be 0.626. The axes are not labelled as the ellipse fitting software uses an RA and Dec-based co-ordinate system which is unphysical for a deprojected image.}
\label{fig:ellipsefit}
\end{figure}

The eccentricity of the disk was constrained by applying an ellipse fit to the outer edge of the deprojected disk image. This was done using an ellipse fitting algorithm that uses edge detection to measure disk profiles. The methods and description of the algorithm will be presented in \cite{Byrne2026} (see figure \ref{fig:ellipsefit}). 

\section{Results}
\label{sec:results}

\subsection{Spiral modelling}
\label{sec:resultsspirals}
\subsubsection{Spiral arm contrast}
\label{sec:resultscontrast}
\begin{figure*}
 \centering
\includegraphics[width=0.95\hsize]{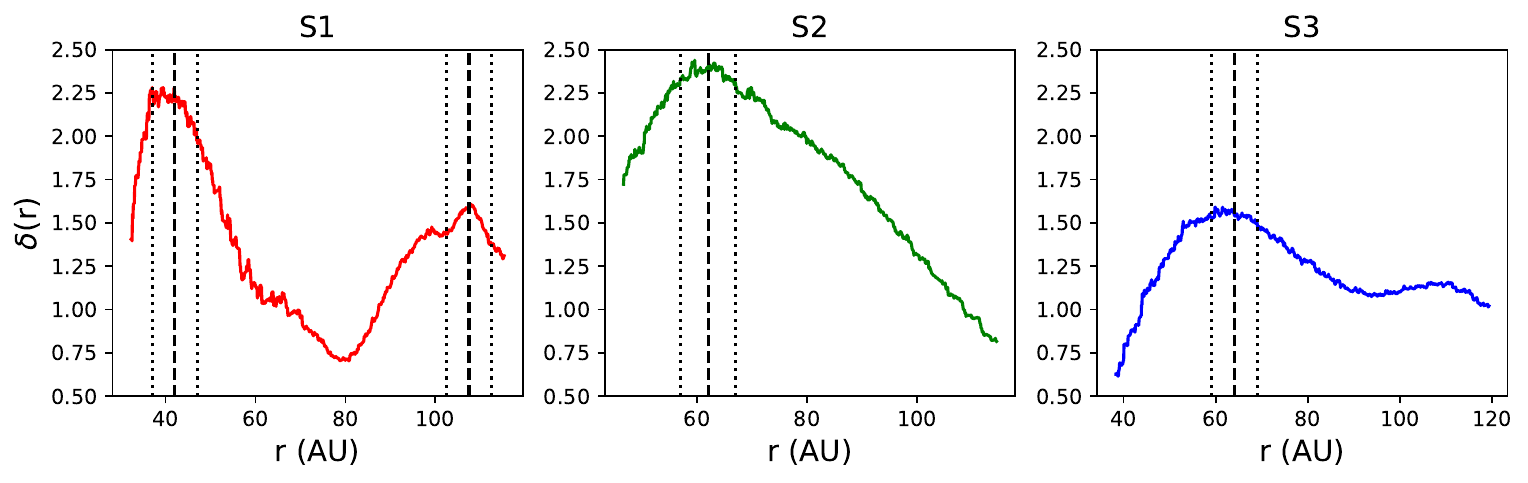}
\caption{Contrast for each observed spiral feature as a function of radius. The black dashed line represents the points at which the spiral intersects the disk major axis. The black dotted lines separated by 4.53 au from the central dashed lines represent the associated spatial error of 1 pixel value.}
\label{fig:spiralcontrast}
\end{figure*}

Figure \ref{fig:spiralcontrast} shows the results of contrast measurement. An important point to note is that the brightness distribution for a polarised disk signal from an inclined disk (see \citealt{benisty2022opticalnearinfraredviewplanetforming} differs from that of the face-on disks simulated in the DF17 models. Additionally the spiral arms seen on a scattering surface represent the peaks of spiral density waves throughout the 3D disk profile and may not represent 2D spiral traces coplanar with the disk midplane. Thus, for a significantly inclined disk the points at which the light scattered by the top surface are most representative of the flux that would be seen from a non-inclined view are the points along the disk major axis; if  the disk were viewed face-on, the brightness distribution across the surface would uniformly match what is seen along the major axis. Thus the most true values for contrast can be found at the regions around which the spirals intersect the major axis, as marked in figure \ref{fig:spiralcontrast} by the black dashed and dotted lines. The peak contrast values lie in this region, and are thus used for comparison with the contrast values found for disk models with $h/r=0.05$ in DF17.

S1 and S2 are in close agreement with the contrast values ranging from $\sim$2.25 to $\sim$2.45, which would suggest a planet mass of the order of 3 Jupiter masses according to the DF17 simulation results. S3 is an outlier here as its contrast value points towards a perturber mass of the order of 1 Jupiter mass, although its extreme opening angle of $\sim85^\circ$ to the tangent of the coronagraph at its starting point should suggest a higher-mass perturber \citep{Zhu_2015}. This extreme launching angle and its consistent modelled pitch angle of $52.5^\circ$ open up the possibility that it is a different kind of feature such as an accretion stream. This deviation in contrast values could mean that (assuming all features are the result of planetary influence) it is less well coupled to the mechanisms driving S1 and S2. We also note that S1 and S2 emerge from behind the coronagraph at very close positions, meaning that they could possibly be a primary and secondary arm in the same spiral system, while within the observing limits imposed by the coronagraph S3 appears to launch from the opposite end of the inner disk. The nature of S3 is further discussed in section \ref{sec:discussion}.

\subsubsection{Spiral arm structure}
\label{sec:resultsarmstructure}

\begin{figure}  
 \centering
\includegraphics[width=\hsize]{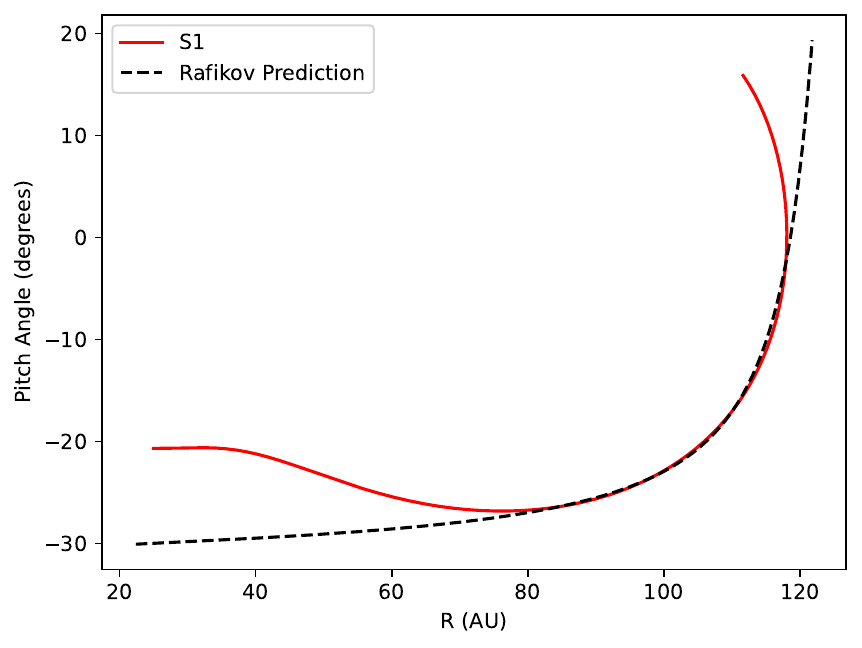}\\
\includegraphics[width=\hsize]{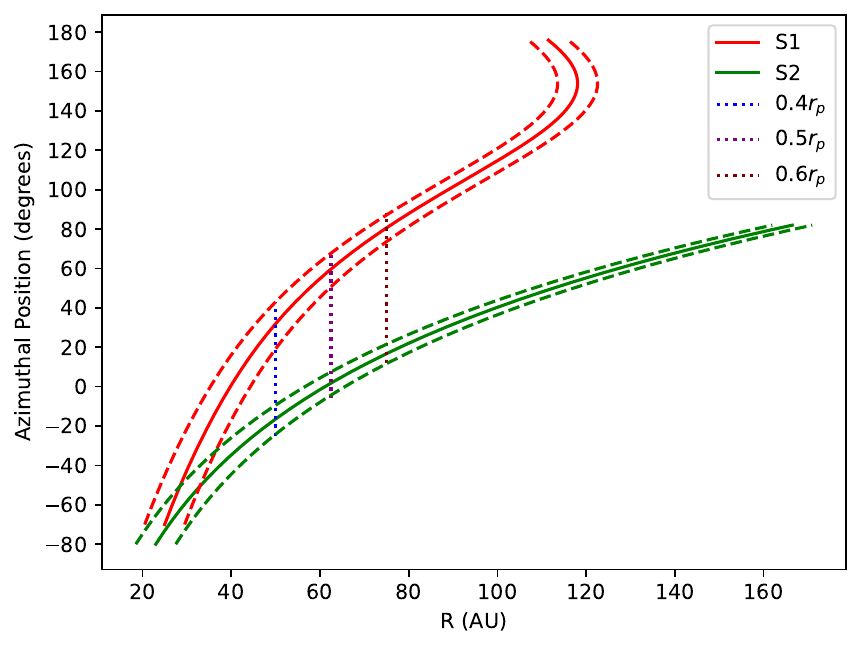}\\
\caption{\textit{Top:} Radially dependent pitch angle for S1 (red line). Negative values tend to the left of the tangent and positive to the right. Overplotted in black is the Rafikov pitch angle model based on equation \ref{eqn:pitchangle} for a planet located at 125 au with $(h/r)_p=0.05$. \textit{Bottom:} Polar co-ordinate plot showing azimuthal angular separation of S1 and S2. Red represents S1, green is S2, the solid line gives the central measured values,  and the dashed lines function as regions covered by error in radial position. The dotted lines mark the angular separations at the positions pertaining to fits for translating separation into mass.}
\label{fig:spiralangles}
\end{figure}

Our results for pitch angle measurement can be seen in figure \ref{fig:spiralangles}. We find that according to linear wave theory S1 best fits a planet position of $125\pm1$ au with $(h/r)_p=0.05\pm0.01$ and $\nu=1/4$ as per the simulations in \cite{Rafikov2002} (see figure \ref{fig:spiralangles}). As S2 and S3 are modelled as logarithmic spirals, they have constant pitch angles that can be found with \begin{equation}\theta=90-\arctan(1/b)\end{equation} to be 35$^{\circ}$ for S2 and 52.5$^{\circ}$ for S3. 

The measurement of azimuthal separation between S1 and S2 is represented in figure \ref{fig:spiralangles} as a polar co-ordinate plot. As the modelling of spiral arms as a function of radius is limited by the resolution of the SPHERE images, radial positions have an associated error of 4.53 au, which is the width of one pixel. To account for this error, each azimuthal separation is measured as a maximum and minimum separation within the margins of error:

for 0.4$r_p$, $30^{\circ}\leq\phi_{p-s}\leq69^{\circ}$ 

$\Rightarrow$ $0.004\text{ M}_{Jup}\leq M_p\leq0.165 \text{ M}_{Jup}$; 

for 0.5$r_p$, $41^{\circ}\leq\phi_{p-s}\leq 74^{\circ}$ 

$\Rightarrow$ $0.01\text{ M}_{Jup}\leq M_p\leq0.22\text{ M}_{Jup}$; 

for 0.6$r_p$, $50^{\circ}\leq \phi_{p-s}=77^{\circ}$ 

$\Rightarrow$ $0.007\text{ M}_{Jup}\leq M_p\leq0.154\text{ M}_{Jup}$. 

With reference to equation \ref{eqn:thermalmass}, the thermal mass for AT Pyx's disk is $\sim $0.08 M$_{Jup}$. Thus, the hypothetical planet mass range found translates to between 0.05 and 2.75$M_{th}$, which is within an acceptable magnitude to excite multiple spiral arms. 
\subsection{Eccentricity}
\label{sec:resultseccentricity}
From this fit an eccentricity of 0.626 was found when measuring the outermost edges of the disk, with the ellipse centre being offset from the star by 0.0724'' or 26.8 au. This measured eccentricity is far in excess of the maximum eccentricity recorded in the simulations of Zhang18 ($\sim$0.3), so this would place a very tentative mass estimate at $\sim3$ M$_{Jup}$, merely because this is the maximum mass probed by the simulations. An inspection of disk ellipse centre offset from star position is slightly more conclusive; plotting the measured offset relative to a planet orbital radius range $100\leq r_p\leq200$ au gives a range for offset/$r_p$ of between 0.13 and 0.27. This puts our planet mass prediction in the range of 3 M$_{Jup}$ with a disk viscosity of $\alpha \approx10^{-3}$ and $h/r = 0.05$ (see figure 8 in \citealt{Zhang_2018}) as these simulations are limited to discrete viscosities, planet masses, and aspect ratios.

\subsection{Summary}
\label{sec:resultssummary}
As a summary of these findings, if all distinct disk characteristics,  i.e. spiral arms and eccentricity,  should be interpreted as signatures of planet formation, then our analyses in these sections would yield a planet mass (assuming a single perturber) between 0.004 and 3 M$_{\text{Jup}}$. This value has an upper limit of 12 M$_{\text{Jup}}$ for a small orbital radius within 0.4'' and 5 M$_{\text{Jup}}$ for an orbital radius outside the main disk complex as within the K-band detection limits any planet above these masses would be visible as a point-source in the K-band image presented in \cite{Ginski_2022}. Additionally, the fitting of a linear wave theory prediction of propagating pitch angle of the most distinct spiral arm estimates an orbital radius of $125\pm1$ au.

It is not possible with this information to conclusively state that planets are present in the disk. Hence other processes giving rise to these signatures are also investigated in the following sections.

\section{Discussion}
\label{sec:discussion}
The exact nature of the feature labelled S3 (figures \ref{fig:bigsphereimage}, \ref{fig:spiralcontrast}) is difficult to determine. Its trajectory, especially in the deprojected image, appears to point radially outward and intersect with the left ansa. In section \ref{sec:spiralarms} it is fitted with a logarithmic spiral with a very large consistent pitch angle, which could effectively be replaced with a straight line in the deprojected image. Its lack of cohesion with spiral features S1 and S2 also limits the value of any of the techniques used in section \ref{sec:spiralarms} for determining its relationship to planet formation. In this section S3 is used as a starting point for investigations into other explanations for disk features.

\subsection{Binarity}
\label{sec:binarity}
One possibility is that S3 is a streamer, a conduit through which material from the outer disk is funnelled towards the inner disk to accrete dust mass onto the star. Such phenomena are known to occur in circumbinary disks such as GG Tau (\citealt{Keppler_2020}, \citealt{Toci2024}), where periodic perturbations within the binary system appear to cause material to be pulled out of the outer disk towards the centre.

Hence, the possibility that AT Pyx is a stellar binary is investigated, first through checking its Gaia renormalised unit weight error (RUWE) value. Given that the presence of a circumstellar disk can inflate a system's RUWE parameter, the threshold for binarity sits at $\sim2.5$ \citep{Fitton_2022}. AT Pyx has a Gaia RUWE value of 3.532 (Gaia DR3, \citealt{DISTANCE_gaia_dr3}), which is a strong indicator of binarity well in excess of the \cite{Fitton_2022} limit of 2.5. It should be considered, however, that Gaia's light measurements can be influenced by AT Pyx's host globule. Similarly to how disk material may present a level of opacity that cannot be fully probed by GAIA \citep{Fitton_2022},  thus introducing more RUWE, a sufficiently bright and dense surrounding cloud could inflate the RUWE parameter. Unfortunately, checking the secular evolution of the light curve from long-term V-band photometric measurements from the ASAS-SN catalogue \citep{Jayasinghe2019} yields no useful information as the timescales for variation are of the order of days with no consistent period; it is a possibility that any brightness variations as a result of a stellar companion may be buried under the variations from non-axisymmetric disk material obscuring the star in a chaotic fashion.

A stellar companion to AT Pyx outside the coronagraph (i.e. outside a radial distance $r=0.0925$''$=34$ au) should be well in excess of the IRDIS K-band lower limit for detection as represented by \citet{Ginski_2022} (see Fig. 5 therein). Additionally the XSHOOTER spectra for AT Pyx should detect lines for a second stellar object to within a separation of 0.1 au for an object with a mass of at least 0.075 M$_\odot$ (using a similar estimation to \citealt{ginski2025diskevolutionstudyimaging}); XSHOOTER is capable of resolving the Doppler-shifted spectral lines of a stellar companion down to an orbital velocity of 16 km/s;  corrected for an inclination of 42.5$^\circ$, this gives a velocity limit of $16/\sin i\approx25$ km/s. The Keplerian orbital velocity of a 0.075 M$_\odot$ companion at 0.1 au is $\sim25$ km/s. This gives a parameter space between 0.1 and 34 au in which an undetected stellar companion could reside. The unobserved parameter space along with the RUWE value mean that the possibility of AT Pyx being a binary system cannot be discounted. When compared with GG Tau \citep{Keppler_2020}, a well-documented circumbinary disk system, S3 has visual parallels with GG Tau's streamers, especially that it could be interpreted as a conduit between the inner and outer disk material given that it intersects the western ansa.

If AT Pyx is a stellar binary, this could also provide an explanation for the disk's eccentricity. Simulations by \cite{Thun2017} and \cite{Ragusa2017} find that the influence of a stellar companion, particularly with higher orbital eccentricity and mass ratio, can drive the opening of an eccentric cavity within the disk. \cite{Ragusa2017} also find that such a process can deposit overdensities of gas and dust in a horseshoe shape at the cavity boundary, with their simulated ALMA continuum observations of such a formation exhibiting similarity to our ALMA 1.3 mm dust continuum image of AT Pyx's disk (see figure \ref{fig:almamoments}).

\subsection{Infall}
\label{sec:infall}

\begin{figure}  
 \centering
\includegraphics[width=0.95\hsize]{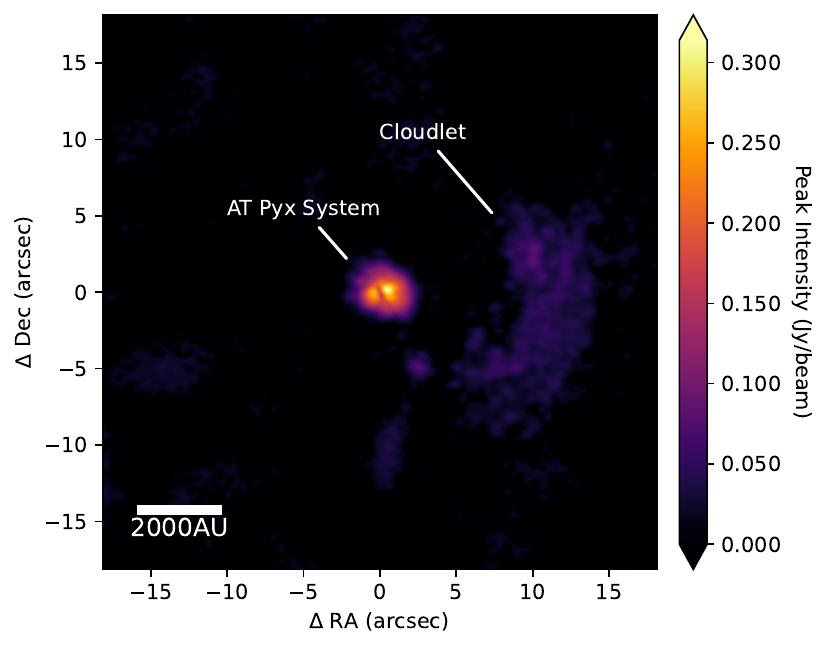}
\includegraphics[width=0.95\hsize]{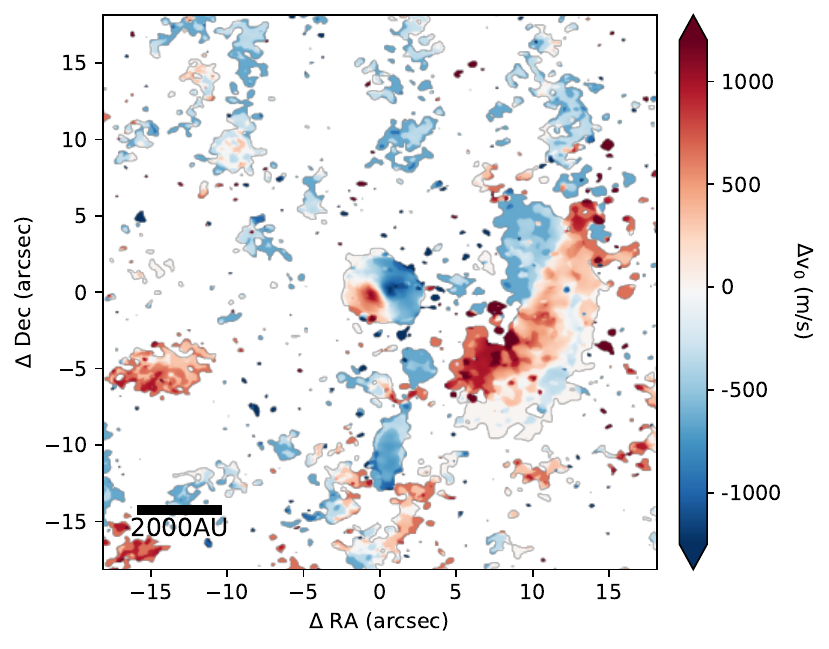}
\caption{\textit{Top:} Moment 8 (peak intensity) image of AT Pyx and nearby cloudlet in \textsuperscript{12}CO gas line emission from ALMA. \textit{Bottom:} Moment 1 (intensity-weighted velocity) image of the same region that demonstrates the velocity components of system and cloudlet.}
\label{fig:environment}
\end{figure}

Streamers are known to be a vital part of mass accretion in the early stages of protostellar evolution (e.g. \citealt{Valdivia-Mena_2024}, \citealt{Cacciapuoti_2023}), but these protostellar streamers do not generally survive past the transition from a protostellar object to a classical T Tauri star, dissipating with the protostellar envelope. However such features can arise visibly in the event of a late-stage infall scenario from substructural cloud material within a young star's surrounding molecular cloud being captured by the star, and in fact their longevity can increase as the star's velocity approaches the systemic velocity of the background gas \citep{Kuffmeier_2019}. With regard to the effects that interplay between star and surrounding cloud can have on disk morphology, there are some qualitative similarities between AT Pyx and simulations of late-stage cloudlet capture, as well as a similar planet--disk system observed to have some continuity between the disk structure and surrounding environment. This section thus investigates both.

A tentative continuity between cloud and disk material can be inferred from an inspection of the kilo-au-scale environment around AT Pyx's disk in \textsuperscript{12}CO emission (see figure \ref{fig:environment}); the large cloud fragment visible on the eastern side of the moment 8 image in particular appears to curl suggestively towards the disk. These formations are not unlike the simulation results presented in \cite{Dullemond_2019}, particularly for the simulation of a small cloud under adiabatic conditions $\sim7700$ years after the first stages of cloud approach (see figure 1 in \citealt{Dullemond_2019}). This may suggest that the surrounding material of CG22 has an ongoing effect on the disk structure.

\begin{figure}  
 \centering
\includegraphics[width=0.85\hsize]{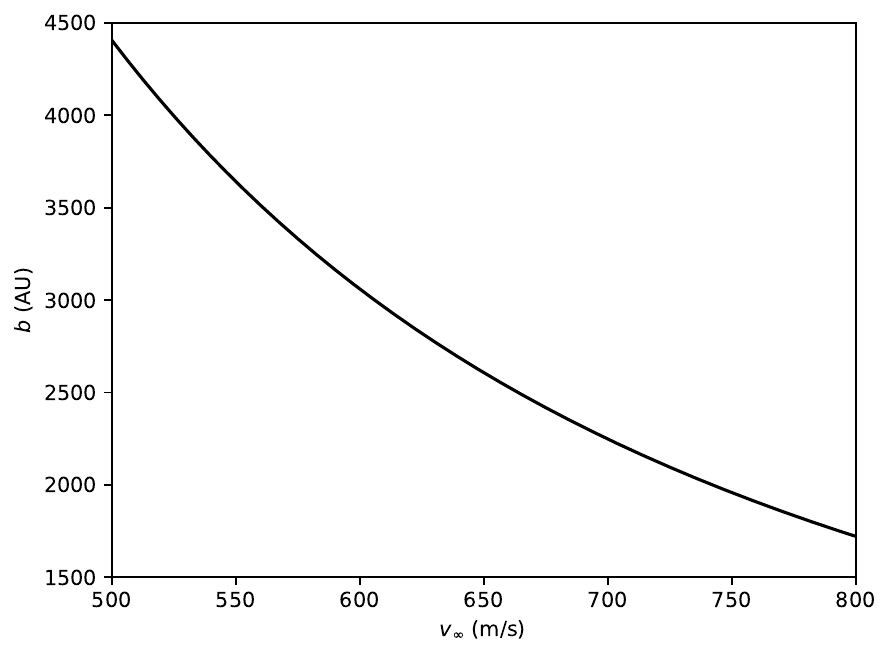}

\caption{Critical impact parameter for AT Pyx as a function of cloud velocity relative to AT Pyx's velocity. This refers to the approach distance between a cloudlet and the AT Pyx system at which the deflection angle of cloud matter around the system is 90$^\circ$.}
\label{fig:impactparam}
\end{figure}

To further investigate the possibility that the cloudlet seen in the kilo-au environment of AT Pyx is infalling, we used a method from \cite{Dullemond_2019}, as follows. The Hoyle-Littleton radius is defined as $R_{HL}=2GM_*/v_\infty^2$, where $v_\infty$ is the cloudlet velocity relative to the star. If the cloudlet size is comparable to $R_{HL}$ then a critical impact parameter 
\begin{equation}
\label{eqn:criticalimpact}
b_{\text{crit}}=\frac{1}{2}R_{HL}=\frac{GM_*}{v_\infty^2}
\end{equation}
can be defined as the approach distance at which the deflection angle of cloud matter around the star is 90$^\circ$, creating a well-defined arc. In the case of the observed cloudlet, the z-velocities are used to give $v_\infty$ ranging between 500 and 800 m/s. From equation \ref{eqn:criticalimpact}, this yields a critical impact parameter within the range 1700-4400 au (see figure \ref{fig:impactparam}). 

While the cloudlet at present is not spherical and these parameters apply broadly to clouds in a pre-capture state and the observed cloudlet can be presumed (if it is interacting with AT Pyx)  to have already undergone capture, a circular region of cloudlet is selected and measured to have a radius of 4.2'' or 1554 au as a lower limit for cloudlet size as it can be assumed that applying this method to the cloud in its current state can provide a qualitative signpost for whether or not it is gravitationally bound to AT Pyx. Because  the cloud fragment radius is below $R_{HL}$ and the cloudlet radial distance of $\sim3220$ au falls within the critical impact parameter range, it can be assumed  that the cloudlet is very likely in some stage of cloud capture.

\begin{figure}  
 \centering
\includegraphics[width=0.95\hsize]{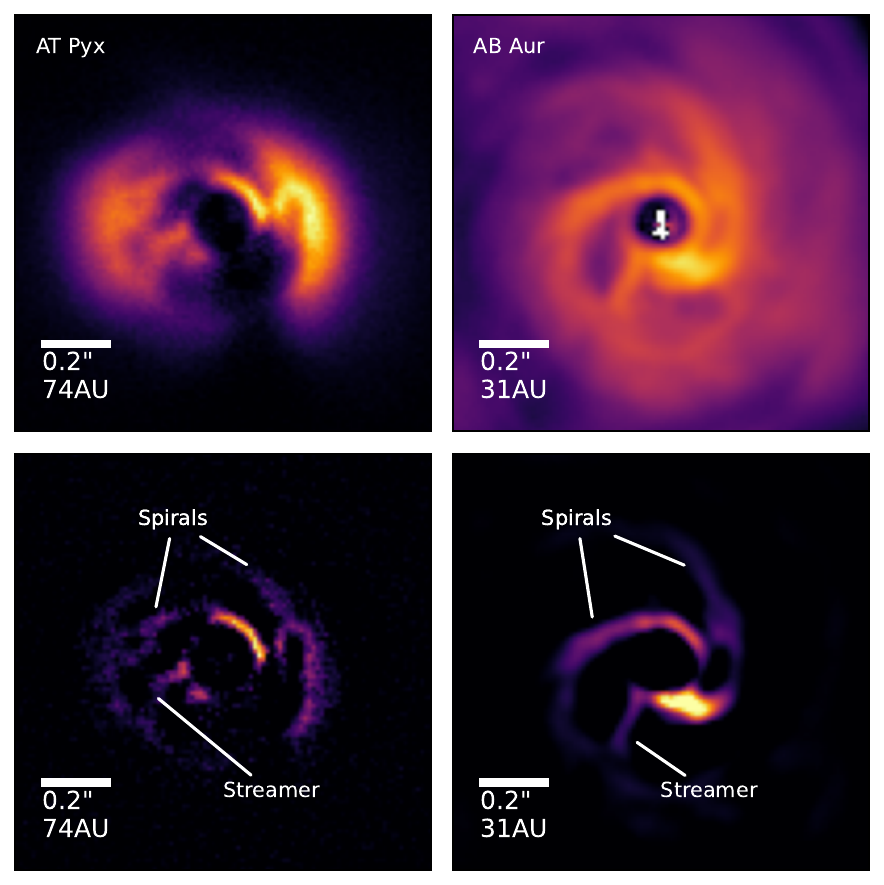}

\caption{Comparison between the spiral morphologies of AT Pyx and AB Aur. AB Aur images have been rotated and flipped horizontally to best demonstrate the structural similarities. \textit{Top:} H-band Q$_{\phi}$ images showing the disk surface in scattered light. \textit{Bottom:} Images through a high-pass filter to better demonstrate spiral structure. We note that that AB Aur's inner spirals bear a striking similarity to AT Pyx's.}
\label{fig:abaurcomp}
\end{figure}

In the case of the comparable system AB Aur, \cite{Speedie_2025} suggest that the disk system is affected by late-stage infall and in its kilo-au-scale surrounding environment it can be seen that a tendril appears to connect the disk with the surrounding cloud structure (figure 8 in \citealt{Speedie_2025}). The infall case for AB Aur could explain the system's formation of spiral features through the infalling material's contribution to the angular momentum of the disk system (\citealt{Speedie_2025}, e.g. \citealt{Lesur_2015}). Presented also in figure \ref{fig:abaurcomp} is a visual comparison between the inner disk regions of AB Aur and AT Pyx in scattered light, where the distribution and nature of spiral arms is uncannily similar, with two arms that appear to trace standard, almost logarithmic, spiral morphologies and one anomalous arm following an almost directly radial path. This would further support the possibility that AT Pyx's features, particularly S3, could be a direct result of an infall scenario. Notably, this hypothesis coexists with the possibility of an embedded planet in AT Pyx's disk, as \cite{Currie2022} suggest the presence of an embedded Jovian protoplanet. To further test the similarities between AT Pyx and AB Aur and the possibility of an infall scenario, we compared their disk masses to a large disk sample, which can be seen in Appendix \ref{sec:dustmasses}.

In terms of non-planet-induced eccentricity the possibility exists that AT Pyx's disk is eccentric as a result of infall. According to simulations by \cite{Kuffmeier_2021}, a secondary circumstellar disk can form after the initial collapse phase due to infall from an encounter between the star and a gas cloudlet, with this secondary disk being eccentric (reaching $e\approx0.4$). It was also shown in these simulations that for a transition disk the inner and outer disks would be misaligned as a result of this infall; misalignment of outer and inner disks can give rise to shadows observed in scattered light from the inner disk casting shadows on the outer disk. Assuming AT Pyx's disk is a late-stage transition disk based on the dip in brightness in the dust continuum (see figure \ref{fig:continuum}) the shadowed region in the south side or far side of the disk in scattered light (see figure \ref{fig:bigsphereimage}) could be interpreted as a result of misalignment between inner and outer disks. 

\section{Conclusions}
\label{sec:conclusions}
This paper presents observations from a range of different instruments and a detailed analysis of AT Pyx's circumstellar disk using data from ESO facilities ALMA, SPHERE, XSHOOTER, and ESPRESSO probing its small dust grain, large dust grain, and gas content, and the central star stellar parameters. Through Keplerian rotation map fitting, photometry, morphological study, image processing, and comparison to theoretical models, this study has led us to draw the following conclusions:

\begin{itemize}
    \item Disk on-sky orientation: The disk is found to have an on-sky position angle of 28.06$\pm0.02$$^{\circ}$ with front scattering and/or near side north of the major axis inclined at an angle of 42.5$\pm0.25^\circ$. This is determined in section \ref{sec:eddy}.
     
    \item Possible origins of the disk's most distinct features, i.e. spirals, rings, eccentricity: The spirals do not appear to be a product of gravitational instability (see Appendix \ref{sec:gravitationalinstability}). Spiral features S1 and S2,  from analyses of contrast, pitch angle, and azimuthal separation in section \ref{sec:spiralarms}, fit the criteria to possibly be a result of planetary influence. Feature S3 is interpreted throughout section \ref{sec:discussion} to possibly be a streamer, either an accretion stream between the inner and outer disks created as a result of perturbations within a central binary's orbit or a streamer resulting from a late infall scenario.
    \item Effect of the Gum Nebula's moderate FUV environment on disk characteristics: AT Pyx's disk's response to its FUV environment is unclear. As mentioned in Appendix \ref{sec:espresso}, no forbidden line tracers strongly indicating external photoevaporation exist within our dataset; however, the possibility of photoevaporation is lent credence by existing observations of 12.81 $\mu$m [Ne\textsc{ii}] line emission from AT Pyx. It is estimated that the FUV field strength at the location of AT Pyx ranges from 0.93 to 27 $G_0$ (see Appendix \ref{sec:fuv}), which is at the lower end of FUV field strengths compared to a survey in Orion by \cite{valegard2024spherevieworionstarforming}. The FUV field's effect on AT Pyx remain inconclusive.
    \item Binarity: AT Pyx has a high Gaia RUWE of 3.532, in excess of the binarity diagnostic limit for disk-hosting systems of 2.5 from \cite{Fitton_2022}. Although it is conceivable that AT Pyx's presence within CG22 has a similar effect to disk material of inflating the RUWE parameter, this can be considered a strong indication of binarity. In Sect. \ref{sec:binarity} we find that there is a parameter space between 0.1 and 34 au within the disk in which a second stellar object could reside. An assumption of binarity could validate S3's status as an accretion stream between the inner and outer disk, much like the streamers observed in GG Tau.
    \item Late-stage infall: From a comparison between AT Pyx's kilo-au environment and simulations in the literature of late-stage cloudlet capture in section \ref{sec:infall}, there is evidence that AT Pyx is currently undergoing late-stage infall, which could be interpreted as a driving force for the disk's spiral features and eccentricity. Additionally, a comparison with the system AB Aur, which is understood as a prominent candidate for late infall, yields striking similarities not only in its kilo-au environment, but also in its inner spiral features, which coexists with the hypothetical presence of a planet as AB Aur is heavily suspected to host a planet itself \citep{Currie2022}.
    \item If a planet has already formed within the disk, the planetary characteristics necessary to drive the observed features: An extensive study into the eccentricity, spiral morphology, and brightness of the observed spiral peaks in sections \ref{sec:spiralarms} and \ref{sec:eccentricity} demonstrates that the comparison of measurements to simulations constrains the possible mass of a single planetary perturber needed to drive the observed features within a range 0.004-3 Jupiter masses. A tentative orbital radius for such a perturber was found from fitting a linear wave theory model to the pitch angle of S1 (the most extended spiral) at $125\pm1$ au. With the current observing capabilities, a planet matching these characteristics cannot be detected. 
\end{itemize}

\begin{acknowledgements}
      We wish to acknowledge the anonymous referee for their helpful comments. J.C.-W. acknowledges funding by the European Union under the Horizon Europe Research \& Innovation Programme 101039452 (WANDA). Views and opinions expressed are, however, those of the author(s) only and do not necessarily reflect those of the European Union or the European Research Council. Neither the European Union nor the granting authority can be held responsible for them. This work makes use of the following ALMA data: ADS/JAO.ALMA\#2021.1.01705.S.  ALMA is a partnership of ESO (representing its member states), NSF (USA) and NINS (Japan), together with NRC (Canada), NSTC and ASIAA (Taiwan), and KASI (Republic of Korea), in cooperation with the Republic of Chile. The Joint ALMA Observatory is operated by ESO, AUI/NRAO and NAOJ. This work has made use of data from the European Space Agency (ESA) mission Gaia (https://www.cosmos.esa.int/gaia), processed by the Gaia Data Processing and Analysis Consortium (DPAC, https://www.cosmos.esa.int/web/gaia/dpac/consortium). Funding for the DPAC has been provided by national institutions, in particular the institutions participating in the Gaia Multilateral Agreement. The author acknowledges the support and supervision provided by the School of Natural Sciences and the Centre for Astronomy at the University of Galway. The author also acknowledges the support and resources provided by the Astronomical Institute and Faculty of Maths and Physics at Charles University. This paper is based on the author’s Master’s thesis carried out at the University of Galway.
\end{acknowledgements}

\bibliographystyle{aa} 
\bibliography{refs}

\begin{thebibliography}{92}
\expandafter\ifx\csname natexlab\endcsname\relax\def\natexlab#1{#1}\fi

\bibitem[{Alexander(2008)}]{Alexander_2008}
Alexander, R.~D. 2008, \mnras: Letters, 391, L64

\bibitem[{{ALMA Partnership} {et~al.}(2015){ALMA Partnership}, Brogan, Pérez, Hunter, Dent, Hales, Hills, Corder, Fomalont, Vlahakis, Asaki, Barkats, Hirota, Hodge, Impellizzeri, Kneissl, Liuzzo, Lucas, Marcelino, Matsushita, Nakanishi, Phillips, Richards, Toledo, Aladro, Broguiere, Cortes, Cortes, Espada, Galarza, Appadoo, Ramirez, Humphreys, Jung, Kameno, Laing, Leon, Marconi, Mignano, Nikolic, Nyman, Radiszcz, Remijan, Rodón, Sawada, Takahashi, Tilanus, Vilaro, Watson, Wiklind, Akiyama, Chapillon, Monsalvo, Francesco, Gueth, Kawamura, Lee, Luong, Mangum, Pietu, Sanhueza, Saigo, Takakuwa, Ubach, Kempen, Wootten, Carrizo, Francke, Gallardo, Garcia, Gonzalez, Hill, Kaminski, Kurono, Liu, Lopez, Morales, Plarre, Schieven, Testi, Videla, Villard, Andreani, Hibbard, \& Tatematsu}]{ALMAPARTNERSHIP_Brogan_2015}
{ALMA Partnership}, Brogan, C.~L., Pérez, L.~M., {et~al.} 2015, \apj, 808, L3

\bibitem[{{Andrews} {et~al.}(2013){Andrews}, {Rosenfeld}, {Kraus}, \& {Wilner}}]{Andrews_2013}
{Andrews}, S.~M., {Rosenfeld}, K.~A., {Kraus}, A.~L., \& {Wilner}, D.~J. 2013, \apj, 771, 129

\bibitem[{{Aru} {et~al.}(2024){Aru}, {Mauc{\'o}}, {Manara}, {Haworth}, {Facchini}, {McLeod}, {Miotello}, {Petr-Gotzens}, {Robberto}, {Rosotti}, {Vicente}, {Winter}, \& {Ansdell}}]{Aru2024}
{Aru}, M.~L., {Mauc{\'o}}, K., {Manara}, C.~F., {et~al.} 2024, \aap, 687, A93

\bibitem[{Avenhaus {et~al.}(2018)Avenhaus, Quanz, Garufi, Perez, Casassus, Pinte, Bertrang, Caceres, Benisty, \& Dominik}]{Avenhaus_2018}
Avenhaus, H., Quanz, S.~P., Garufi, A., {et~al.} 2018, \apj, 863, 44

\bibitem[{Bae \& Zhu(2018)}]{Bae_2018b}
Bae, J. \& Zhu, Z. 2018, \apj, 859, 119

\bibitem[{Baehr \& Zhu(2021)}]{Baehr_2021}
Baehr, H. \& Zhu, Z. 2021, \apj, 909, 135

\bibitem[{{Ballabio} {et~al.}(2023){Ballabio}, {Haworth}, \& {Henney}}]{Ballabio2023}
{Ballabio}, G., {Haworth}, T.~J., \& {Henney}, W.~J. 2023, \mnras, 518, 5563

\bibitem[{{Baraffe} {et~al.}(2015){Baraffe}, {Homeier}, {Allard}, \& {Chabrier}}]{baraffe2015}
{Baraffe}, I., {Homeier}, D., {Allard}, F., \& {Chabrier}, G. 2015, A\&A, 577, A42

\bibitem[{{Benisty} {et~al.}(2023){Benisty}, {Dominik}, {Follette}, {Garufi}, {Ginski}, {Hashimoto}, {Keppler}, {Kley}, \& {Monnier}}]{benisty2022opticalnearinfraredviewplanetforming}
{Benisty}, M., {Dominik}, C., {Follette}, K., {et~al.} 2023, in ASPC, Vol. 534, Protostars and Planets VII, ed. S.~{Inutsuka}, Y.~{Aikawa}, T.~{Muto}, K.~{Tomida}, \& M.~{Tamura}, 605

\bibitem[{{Bertoldi}(1989)}]{Bertoldi1989B}
{Bertoldi}, F. 1989, \apj, 346, 735

\bibitem[{{Bok} \& {Reilly}(1947)}]{BokandReilly1947}
{Bok}, B.~J. \& {Reilly}, E.~F. 1947, \apj, 105, 255

\bibitem[{Bradley {et~al.}(2025)Bradley, Sipőcz, Robitaille, Tollerud, Vinícius, Deil, Barbary, Wilson, Busko, Donath, Günther, Cara, Lim, Meßlinger, Burnett, Conseil, Droettboom, Bostroem, Bray, Bratholm, Jamieson, Ginsburg, Barentsen, Craig, Pascual, Rathi, Perrin, \& Morris}]{larry_bradley_photutils_2025_14889440}
Bradley, L., Sipőcz, B., Robitaille, T., {et~al.} 2025, astropy/photutils: 2.2.0

\bibitem[{{Byrne} {et~al.}(2026){Byrne}, {Ginski}, {van Capelleveen}, {Fitzgerald}, {Garufi}, {Lawlor}, \& {McLachlan}}]{Byrne2026}
{Byrne}, J., {Ginski}, C., {van Capelleveen}, R., {et~al.} 2026, \aap, submitted

\bibitem[{{Cacciapuoti} {et~al.}(2024){Cacciapuoti}, {Macias}, {Gupta}, {Testi}, {Miotello}, {Espaillat}, {K{\"u}ffmeier}, {van Terwisga}, {Tobin}, {Grant}, {Manara}, {Segura-Cox}, {Wendeborn}, {Klessen}, {Maury}, {Lebreuilly}, {Hennebelle}, \& {Molinari}}]{Cacciapuoti_2023}
{Cacciapuoti}, L., {Macias}, E., {Gupta}, A., {et~al.} 2024, \aap, 682, A61

\bibitem[{{Campbell-White} {et~al.}(2021){Campbell-White}, {Sicilia-Aguilar}, {Manara}, {Matsumura}, {Fang}, {Frasca}, \& {Roccatagliata}}]{Campbellwhite2021}
{Campbell-White}, J., {Sicilia-Aguilar}, A., {Manara}, C.~F., {et~al.} 2021, \mnras, 507, 3331

\bibitem[{{Caraveo} {et~al.}(2001){Caraveo}, {De Luca}, {Mignani}, \& {Bignami}}]{Caraveo2001velapulsardist}
{Caraveo}, P.~A., {De Luca}, A., {Mignani}, R.~P., \& {Bignami}, G.~F. 2001, \apj, 561, 930

\bibitem[{{Chanot} \& {Sivan}(1983)}]{ChanotandSivan1983}
{Chanot}, A. \& {Sivan}, J.~P. 1983, \aap, 121, 19

\bibitem[{{Chiang} {et~al.}(2001){Chiang}, {Joung}, {Creech-Eakman}, {Qi}, {Kessler}, {Blake}, \& {van Dishoeck}}]{Chiang2001}
{Chiang}, E.~I., {Joung}, M.~K., {Creech-Eakman}, M.~J., {et~al.} 2001, \apj, 547, 1077

\bibitem[{Choudhury \& Bhatt(2009)}]{Choudhury2009GumNebulaGlobules}
Choudhury, R. \& Bhatt, H.~C. 2009, \mnras, 393, 959

\bibitem[{{Claes} {et~al.}(2024){Claes}, {Campbell-White}, {Manara}, {Frasca}, {Natta}, {Alcal{\'a}}, {Armeni}, {Fang}, {Lovell}, {Stelzer}, {Venuti}, {Wyatt}, \& {Queitsch}}]{Claes2024}
{Claes}, R.~A.~B., {Campbell-White}, J., {Manara}, C.~F., {et~al.} 2024, \aap, 690, A122

\bibitem[{Cossins {et~al.}(2009)Cossins, Lodato, \& Clarke}]{Cossins_2009}
Cossins, P., Lodato, G., \& Clarke, C.~J. 2009, \mnras, 393, 1157–1173

\bibitem[{{Currie} {et~al.}(2022){Currie}, {Lawson}, {Schneider}, {Lyra}, {Wisniewski}, {Grady}, {Guyon}, {Tamura}, {Kotani}, {Kawahara}, {Brandt}, {Uyama}, {Muto}, {Dong}, {Kudo}, {Hashimoto}, {Fukagawa}, {Wagner}, {Lozi}, {Chilcote}, {Tobin}, {Groff}, {Ward-Duong}, {Januszewski}, {Norris}, {Tuthill}, {van der Marel}, {Sitko}, {Deo}, {Vievard}, {Jovanovic}, {Martinache}, \& {Skaf}}]{Currie2022}
{Currie}, T., {Lawson}, K., {Schneider}, G., {et~al.} 2022, \nat, 6, 751

\bibitem[{{Dong} \& {Fung}(2017)}]{Dong_2017}
{Dong}, R. \& {Fung}, J. 2017, \apj, 835, 38

\bibitem[{Dong {et~al.}(2016)Dong, Vorobyov, Pavlyuchenkov, Chiang, \& Liu}]{Dong_2016SpiralInstability}
Dong, R., Vorobyov, E., Pavlyuchenkov, Y., Chiang, E., \& Liu, H.~B. 2016, \apj, 823, 141

\bibitem[{{Dullemond} {et~al.}(2019){Dullemond}, {K{\"u}ffmeier}, {Goicovic}, {Fukagawa}, {Oehl}, \& {Kramer}}]{Dullemond_2019}
{Dullemond}, C.~P., {K{\"u}ffmeier}, M., {Goicovic}, F., {et~al.} 2019, \aap, 628, A20

\bibitem[{{Durisen} {et~al.}(2007){Durisen}, {Boss}, {Mayer}, {Nelson}, {Quinn}, \& {Rice}}]{Durisen2007}
{Durisen}, R.~H., {Boss}, A.~P., {Mayer}, L., {et~al.} 2007, in Protostars and Planets V, ed. B.~{Reipurth}, D.~{Jewitt}, \& K.~{Keil}, 607

\bibitem[{{Elmegreen}(1998)}]{elmegreen1998observationstheorydynamicaltriggers}
{Elmegreen}, B.~G. 1998, in ASPC, Vol. 148, Origins, ed. C.~E. {Woodward}, J.~M. {Shull}, \& H.~A. {Thronson}, Jr., 150

\bibitem[{{Ercolano} \& {Owen}(2010)}]{Ercolano2010}
{Ercolano}, B. \& {Owen}, J.~E. 2010, \mnras, 406, 1553

\bibitem[{{Ercolano} \& {Picogna}(2022)}]{Ercolano2022}
{Ercolano}, B. \& {Picogna}, G. 2022, European Physical Journal Plus, 137, 1357

\bibitem[{Fitton {et~al.}(2022)Fitton, Tofflemire, \& Kraus}]{Fitton_2022}
Fitton, S., Tofflemire, B.~M., \& Kraus, A.~L. 2022, RNAAS, 6, 18

\bibitem[{Fung \& Dong(2015)}]{Fung__2015spiralangleseparation}
Fung, J. \& Dong, R. 2015, \apj, 815, L21

\bibitem[{{Gaia Collaboration} {et~al.}(2021){Gaia Collaboration}, {Vallenari}, {Prusti}, {de Bruijne}, {Babusiaux}, \& et~al.}]{DISTANCE_gaia_dr3}
{Gaia Collaboration}, {Brown}, A.~G.~A., {Vallenari}, A., {Prusti}, T., {et~al.} 2021, A\&A, 649, A1

\bibitem[{{Garufi} {et~al.}(2024){Garufi}, {Ginski}, {van Holstein}, {Benisty}, {Manara}, {P{\'e}rez}, {Pinilla}, {Ribas}, {Weber}, {Williams}, {Cieza}, {Dominik}, {Facchini}, {Huang}, {Zblaho}, {Bae}, {Hagelberg}, {Henning}, {Hogerheijde}, {Janson}, {M{\'e}nard}, {Messina}, {Meyer}, {Pinte}, {Quanz}, {Rigliaco}, {Roccatagliata}, {Schmid}, {Szul{\'a}gyi}, {van Boekel}, {Wahhaj}, {Antichi}, {Baruffolo}, \& {Moulin}}]{Garufi2024}
{Garufi}, A., {Ginski}, C., {van Holstein}, R.~G., {et~al.} 2024, \aap, 685, A53

\bibitem[{{Ginski} {et~al.}(2024){Ginski}, {Garufi}, {Benisty}, {Tazaki}, {Dominik}, {Ribas}, {Engler}, {Birnstiel}, {Chauvin}, {Columba}, {Facchini}, {Goncharov}, {Hagelberg}, {Henning}, {Hogerheijde}, {van Holstein}, {Huang}, {Muto}, {Pinilla}, {Kanagawa}, {Kim}, {Kurtovic}, {Langlois}, {Manara}, {Milli}, {Momose}, {Orihara}, {Pawellek}, {Pinte}, {Rab}, {Schmidt}, {Snik}, {Wahhaj}, {Williams}, \& {Zblaho}}]{Ginski2024Chamaeleon}
{Ginski}, C., {Garufi}, A., {Benisty}, M., {et~al.} 2024, \aap, 685, A52

\bibitem[{Ginski {et~al.}(2022)Ginski, Gratton, Bohn, Dominik, Jorquera, Chauvin, Milli, Rodriguez, Benisty, Launhardt, Müller, Cugno, van Holstein, Boccaletti, Muro-Arena, Desidera, Keppler, Zblaho, Sissa, Henning, Janson, Langlois, Bonnefoy, Cantalloube, D’Orazi, Feldt, Hagelberg, Ségransan, Lagrange, Lazzoni, Meyer, Romero, Schmidt, Vigan, Petit, Roelfsema, Pragt, \& Weber}]{Ginski_2022}
Ginski, C., Gratton, R., Bohn, A., {et~al.} 2022, \aap, 662, A74

\bibitem[{{Ginski} {et~al.}(2025){Ginski}, {Pinilla}, {Benisty}, {Pinte}, {Claes}, {Mamajek}, {Kenworthy}, {Murphy}, {Manara}, {Bae}, {Birnstiel}, {Byrne}, {Dominik}, {Facchini}, {Garufi}, {Gratton}, {Hogerheijde}, {van Holstein}, {Huang}, {Langlois}, {Lawlor}, {Ma}, {McLachlan}, {Menard}, {Rigliaco}, {Ribas}, {Schmidt}, {Sierra}, {Tazaki}, {Williams}, \& {Zurlo}}]{ginski2025diskevolutionstudyimaging}
{Ginski}, C., {Pinilla}, P., {Benisty}, M., {et~al.} 2025, \aap, 699, A237

\bibitem[{{Glassgold} {et~al.}(2007){Glassgold}, {Najita}, \& {Igea}}]{Glassgold2007}
{Glassgold}, A.~E., {Najita}, J.~R., \& {Igea}, J. 2007, \apj, 656, 515

\bibitem[{{Goldreich} \& {Tremaine}(1979)}]{GoldreichandTremaineLinearWaveTheory1979}
{Goldreich}, P. \& {Tremaine}, S. 1979, \apj, 233, 857

\bibitem[{{Goodman} \& {Rafikov}(2001)}]{GoodmanRafikov2001}
{Goodman}, J. \& {Rafikov}, R.~R. 2001, \apj, 552, 793

\bibitem[{{G{\"u}del} {et~al.}(2010){G{\"u}del}, {Lahuis}, {Briggs}, {Carr}, {Glassgold}, {Henning}, {Najita}, {van Boekel}, \& {van Dishoeck}}]{Gudel2010}
{G{\"u}del}, M., {Lahuis}, F., {Briggs}, K.~R., {et~al.} 2010, \aap, 519, A113

\bibitem[{{Gum}(1952)}]{Gum1952}
{Gum}, C.~S. 1952, The Observatory, 72, 151

\bibitem[{{Habing}(1968)}]{Habing1968}
{Habing}, H.~J. 1968, \bain, 19, 421

\bibitem[{Hartmann(2008)}]{Hartmann_2008}
Hartmann, L. 2008, Physica Scripta, 2008, 014012

\bibitem[{{Hawarden} \& {Brand}(1976)}]{HawardenandBrand1976}
{Hawarden}, T.~G. \& {Brand}, P.~W.~J.~L. 1976, \mnras, 175, 19P

\bibitem[{{Herczeg} \& {Hillenbrand}(2014)}]{HerczegandHillenbrand2014}
{Herczeg}, G.~J. \& {Hillenbrand}, L.~A. 2014, \apj, 786, 97

\bibitem[{{Hildebrand}(1983)}]{Hildebrand_1983}
{Hildebrand}, R.~H. 1983, \qjras, 24, 267

\bibitem[{Isella {et~al.}(2016)Isella, Guidi, Testi, Liu, Li, Li, Weaver, Boehler, Carperter, De~Gregorio-Monsalvo, Manara, Natta, P\'erez, Ricci, Sargent, Tazzari, \& Turner}]{Isella2016}
Isella, A., Guidi, G., Testi, L., {et~al.} 2016, Phys. Rev. Lett., 117, 251101

\bibitem[{{Jayasinghe} {et~al.}(2019){Jayasinghe}, {Stanek}, {Kochanek}, {Shappee}, {Holoien}, {Thompson}, {Prieto}, {Dong}, {Pawlak}, {Pejcha}, {Shields}, {Pojmanski}, {Otero}, {Britt}, \& {Will}}]{Jayasinghe2019}
{Jayasinghe}, T., {Stanek}, K.~Z., {Kochanek}, C.~S., {et~al.} 2019, \mnras, 486, 1907

\bibitem[{Keppler {et~al.}(2020)Keppler, Penzlin, Benisty, van Boekel, Henning, van Holstein, Kley, Garufi, Ginski, Brandner, Bertrang, Boccaletti, de~Boer, Bonavita, Brown~Sevilla, Chauvin, Dominik, Janson, Langlois, Lodato, Maire, Ménard, Pantin, Pinte, Stolker, Szulágyi, Thebault, Villenave, Zblaho, Rabou, Feautrier, Feldt, Madec, \& Wildi}]{Keppler_2020}
Keppler, M., Penzlin, A., Benisty, M., {et~al.} 2020, \aap, 639, A62

\bibitem[{Kim {et~al.}(2005)Kim, Walter, \& Wolk}]{Kim_2005}
Kim, J.~S., Walter, F.~M., \& Wolk, S.~J. 2005, \apj, 129, 1564

\bibitem[{{Kley} \& {Dirksen}(2006)}]{kley2005}
{Kley}, W. \& {Dirksen}, G. 2006, A\&A, 447, 369

\bibitem[{{Kristensen} {et~al.}(2012){Kristensen}, {van Dishoeck}, {Bergin}, {Visser}, {Y{\i}ld{\i}z}, {San Jose-Garcia}, {J{\o}rgensen}, {Herczeg}, {Johnstone}, {Wampfler}, {Benz}, {Bruderer}, {Cabrit}, {Caselli}, {Doty}, {Harsono}, {Herpin}, {Hogerheijde}, {Karska}, {van Kempen}, {Liseau}, {Nisini}, {Tafalla}, {van der Tak}, \& {Wyrowski}}]{Kristensen_2012}
{Kristensen}, L.~E., {van Dishoeck}, E.~F., {Bergin}, E.~A., {et~al.} 2012, \aap, 542, A8

\bibitem[{Kuffmeier {et~al.}(2021)Kuffmeier, Dullemond, Reissl, \& Goicovic}]{Kuffmeier_2021}
Kuffmeier, M., Dullemond, C.~P., Reissl, S., \& Goicovic, F.~G. 2021, \aap, 656, A161

\bibitem[{Kuffmeier {et~al.}(2019)Kuffmeier, Goicovic, \& Dullemond}]{Kuffmeier_2019}
Kuffmeier, M., Goicovic, F.~G., \& Dullemond, C.~P. 2019, \aap, 633, A3

\bibitem[{{Lesur} {et~al.}(2015){Lesur}, {Hennebelle}, \& {Fromang}}]{Lesur_2015}
{Lesur}, G., {Hennebelle}, P., \& {Fromang}, S. 2015, \aap, 582, L9

\bibitem[{{Lin} \& {Shu}(1964)}]{Lin1964}
{Lin}, C.~C. \& {Shu}, F.~H. 1964, \apj, 140, 646

\bibitem[{{Lin} \& {Papaloizou}(1993)}]{LinPapaloizou1993}
{Lin}, D.~N.~C. \& {Papaloizou}, J.~C.~B. 1993, in Protostars and Planets III, ed. E.~H. {Levy} \& J.~I. {Lunine}, 749

\bibitem[{{Ma{\'\i}z Apell{\'a}niz} {et~al.}(2008){Ma{\'\i}z Apell{\'a}niz}, {Alfaro}, \& {Sota}}]{apellániz2008}
{Ma{\'\i}z Apell{\'a}niz}, J., {Alfaro}, E.~J., \& {Sota}, A. 2008, arXiv e-prints, arXiv:0804.2553

\bibitem[{{Manara} {et~al.}(2023){Manara}, {Ansdell}, {Rosotti}, {Hughes}, {Armitage}, {Lodato}, \& {Williams}}]{manara2023}
{Manara}, C.~F., {Ansdell}, M., {Rosotti}, G.~P., {et~al.} 2023, in ASPC, Vol. 534, Protostars and Planets VII, ed. S.~{Inutsuka}, Y.~{Aikawa}, T.~{Muto}, K.~{Tomida}, \& M.~{Tamura}, 539

\bibitem[{{Muto} {et~al.}(2012){Muto}, {Grady}, {Hashimoto}, {Fukagawa}, {Hornbeck}, {Sitko}, {Russell}, {Werren}, {Cur{\'e}}, {Currie}, {Ohashi}, {Okamoto}, {Momose}, {Honda}, {Inutsuka}, {Takeuchi}, {Dong}, {Abe}, {Brandner}, {Brandt}, {Carson}, {Egner}, {Feldt}, {Fukue}, {Goto}, {Guyon}, {Hayano}, {Hayashi}, {Hayashi}, {Henning}, {Hodapp}, {Ishii}, {Iye}, {Janson}, {Kandori}, {Knapp}, {Kudo}, {Kusakabe}, {Kuzuhara}, {Matsuo}, {Mayama}, {McElwain}, {Miyama}, {Morino}, {Moro-Martin}, {Nishimura}, {Pyo}, {Serabyn}, {Suto}, {Suzuki}, {Takami}, {Takato}, {Terada}, {Thalmann}, {Tomono}, {Turner}, {Watanabe}, {Wisniewski}, {Yamada}, {Takami}, {Usuda}, \& {Tamura}}]{Muto2012}
{Muto}, T., {Grady}, C.~A., {Hashimoto}, J., {et~al.} 2012, \apjl, 748, L22

\bibitem[{{O'Dell} \& {Wen}(1994)}]{Odell1994}
{O'Dell}, C.~R. \& {Wen}, Z. 1994, \apj, 436, 194

\bibitem[{Ogilvie \& Lubow(2002)}]{Ogilvie_2002}
Ogilvie, G.~I. \& Lubow, S.~H. 2002, \mnras, 330, 950–954

\bibitem[{Pascucci {et~al.}(2007)Pascucci, Hollenbach, Najita, Muzerolle, Gorti, Herczeg, Hillenbrand, Kim, Carpenter, Meyer, Mamajek, \& Bouwman}]{Pascucci_2007}
Pascucci, I., Hollenbach, D., Najita, J., {et~al.} 2007, \apj, 663, 383–393

\bibitem[{Pecaut \& Mamajek(2013)}]{Pecaut_2013}
Pecaut, M.~J. \& Mamajek, E.~E. 2013, \apjs, 208, 9

\bibitem[{Pozzo {et~al.}(2000)Pozzo, Jeffries, Naylor, Totten, Harmer, \& Kenyon}]{Pozzo2000}
Pozzo, M., Jeffries, R.~D., Naylor, T., {et~al.} 2000, \mnras, 313, L23

\bibitem[{{Rafikov}(2002)}]{Rafikov2002}
{Rafikov}, R.~R. 2002, \apj, 569, 997

\bibitem[{{Ragusa} {et~al.}(2017){Ragusa}, {Dipierro}, {Lodato}, {Laibe}, \& {Price}}]{Ragusa2017}
{Ragusa}, E., {Dipierro}, G., {Lodato}, G., {Laibe}, G., \& {Price}, D.~J. 2017, \mnras, 464, 1449

\bibitem[{Reggiani {et~al.}(2018)Reggiani, Christiaens, Absil, Mawet, Huby, Choquet, Gomez~Gonzalez, Ruane, Femenia, Serabyn, Matthews, Barraza, Carlomagno, Defrère, Delacroix, Habraken, Jolivet, Karlsson, Orban~de Xivry, Piron, Surdej, Vargas~Catalan, \& Wertz}]{Reggiani_2018}
Reggiani, M., Christiaens, V., Absil, O., {et~al.} 2018, \aap, 611, A74

\bibitem[{{Reipurth}(1983)}]{Reipurth1983}
{Reipurth}, B. 1983, \aap, 117, 183

\bibitem[{{Rich} {et~al.}(2022){Rich}, {Monnier}, {Aarnio}, {Laws}, {Setterholm}, {Wilner}, {Calvet}, {Harries}, {Miller}, {Davies}, {Adams}, {Andrews}, {Bae}, {Espaillat}, {Greenbaum}, {Hinkley}, {Kraus}, {Hartmann}, {Isella}, {McClure}, {Oppenheimer}, {P{\'e}rez}, \& {Zhu}}]{Rich2022}
{Rich}, E.~A., {Monnier}, J.~D., {Aarnio}, A., {et~al.} 2022, \aj, 164, 109

\bibitem[{{Rivi{\`e}re-Marichalar} {et~al.}(2024){Rivi{\`e}re-Marichalar}, {Mac{\'\i}as}, {Baruteau}, {Fuente}, {Neri}, {Ribas}, {Esplugues}, {Navarro-Almaida}, {Osorio}, \& {Anglada}}]{rivièremarichalar2023}
{Rivi{\`e}re-Marichalar}, P., {Mac{\'\i}as}, E., {Baruteau}, C., {et~al.} 2024, \aap, 683, A141

\bibitem[{{Sahu} \& {Sahu}(1992)}]{SahuandSahu1992}
{Sahu}, M. \& {Sahu}, K.~C. 1992, \aap, 259, 265

\bibitem[{Sheehan {et~al.}(2022)Sheehan, Tobin, Li, van~’t Hoff, Jørgensen, Kwon, Looney, Ohashi, Takakuwa, Williams, Aso, Gavino, Gregorio-Monsalvo, Han, Lee, Plunkett, Sharma, Aikawa, Lai, Lee, Lin, Saigo, Tomida, \& Yen}]{Sheehan_2022}
Sheehan, P.~D., Tobin, J.~J., Li, Z.-Y., {et~al.} 2022, \apj, 934, 95

\bibitem[{Speedie {et~al.}(2025)Speedie, Dong, Teague, Segura-Cox, Pineda, Calcino, Longarini, Hall, Tang, Hashimoto, Paneque-Carreño, Lodato, \& Veronesi}]{Speedie_2025}
Speedie, J., Dong, R., Teague, R., {et~al.} 2025, \apj, 981, L30

\bibitem[{Takeuchi \& Artymowicz(2001)}]{Takeuchi_Artymowicz_2001}
Takeuchi, T. \& Artymowicz, P. 2001, \apj, 557, 990–1006

\bibitem[{Teague(2019)}]{eddy}
Teague, R. 2019, JOSS, 4, 1220

\bibitem[{Teague {et~al.}(2019)Teague, Bae, Huang, \& Bergin}]{Teague_2019}
Teague, R., Bae, J., Huang, J., \& Bergin, E.~A. 2019, \apj, 884, L56

\bibitem[{{Teague} \& {Foreman-Mackey}(2018)}]{bettermoments}
{Teague}, R. \& {Foreman-Mackey}, D. 2018, RNAAS, 2, 173

\bibitem[{{Thun} {et~al.}(2017){Thun}, {Kley}, \& {Picogna}}]{Thun2017}
{Thun}, D., {Kley}, W., \& {Picogna}, G. 2017, \aap, 604, A102

\bibitem[{{Toci} {et~al.}(2024){Toci}, {Ceppi}, {Cuello}, {Duch{\^e}ne}, {Ragusa}, {Lodato}, {Farina}, {M{\'e}nard}, \& {Aly}}]{Toci2024}
{Toci}, C., {Ceppi}, S., {Cuello}, N., {et~al.} 2024, \aap, 688, A102

\bibitem[{{Toomre}(1964)}]{Toomre1964}
{Toomre}, A. 1964, \apj, 139, 1217

\bibitem[{{Valdivia-Mena} {et~al.}(2024){Valdivia-Mena}, {Pineda}, {Caselli}, {Segura-Cox}, {Schmiedeke}, {Spezzano}, {Offner}, {Ivlev}, {Kuffmeier}, {Cunningham}, {Neri}, \& {Maureira}}]{Valdivia-Mena_2024}
{Valdivia-Mena}, M.~T., {Pineda}, J.~E., {Caselli}, P., {et~al.} 2024, \aap, 687, A71

\bibitem[{{Valeg{\r{a}}rd} {et~al.}(2024){Valeg{\r{a}}rd}, {Ginski}, {Derkink}, {Garufi}, {Dominik}, {Ribas}, {Williams}, {Benisty}, {Birnstiel}, {Facchini}, {Columba}, {Hogerheijde}, {van Holstein}, {Huang}, {Kenworthy}, {Manara}, {Pinilla}, {Rab}, {Sulaiman}, \& {Zurlo}}]{valegard2024spherevieworionstarforming}
{Valeg{\r{a}}rd}, P.~G., {Ginski}, C., {Derkink}, A., {et~al.} 2024, \aap, 685, A54

\bibitem[{{van Capelleveen} {et~al.}(2025){van Capelleveen}, {Ginski}, {Kenworthy}, {Byrne}, {Lawlor}, {McLachlan}, {Mamajek}, {Stolker}, {Benisty}, {Bohn}, {Close}, {Dominik}, {Haffert}, {Landman}, {Ma}, {Snellen}, {Tazaki}, {van der Marel}, {Welzel}, \& {Zhang}}]{vancapelleveen2025}
{van Capelleveen}, R.~F., {Ginski}, C., {Kenworthy}, M.~A., {et~al.} 2025, \apjl, 990, L8

\bibitem[{{van Holstein} {et~al.}(2020){van Holstein}, {Girard}, {de Boer}, {Snik}, {Milli}, {Stam}, {Ginski}, {Mouillet}, {Wahhaj}, {Schmid}, {Keller}, {Langlois}, {Dohlen}, {Vigan}, {Pohl}, {Carbillet}, {Fantinel}, {Maurel}, {Orign{\'e}}, {Petit}, {Ramos}, {Rigal}, {Sevin}, {Boccaletti}, {Le Coroller}, {Dominik}, {Henning}, {Lagadec}, {M{\'e}nard}, {Turatto}, {Udry}, {Chauvin}, {Feldt}, \& {Beuzit}}]{irdapvanholstein_2020}
{van Holstein}, R.~G., {Girard}, J.~H., {de Boer}, J., {et~al.} 2020, \aap, 633, A64

\bibitem[{{Weber} {et~al.}(2020){Weber}, {Ercolano}, {Picogna}, {Hartmann}, \& {Rodenkirch}}]{Weber2020diskwind}
{Weber}, M.~L., {Ercolano}, B., {Picogna}, G., {Hartmann}, L., \& {Rodenkirch}, P.~J. 2020, \mnras, 496, 223

\bibitem[{Williams \& Cieza(2011)}]{Williams_2011}
Williams, J.~P. \& Cieza, L.~A. 2011, ARA\&A, 49, 67–117

\bibitem[{Winter {et~al.}(2018)Winter, Clarke, Rosotti, Ih, Facchini, \& Haworth}]{Winter2018}
Winter, A.~J., Clarke, C.~J., Rosotti, G., {et~al.} 2018, \mnras, 478, 2700

\bibitem[{{Yep} \& {White}(2020)}]{Yep2020}
{Yep}, A.~C. \& {White}, R.~J. 2020, \apj, 889, 50

\bibitem[{Zhang {et~al.}(2018)Zhang, Zhu, Huang, Guzmán, Andrews, Birnstiel, Dullemond, Carpenter, Isella, Pérez, Benisty, Wilner, Baruteau, Bai, \& Ricci}]{Zhang_2018}
Zhang, S., Zhu, Z., Huang, J., {et~al.} 2018, \apj, 869, L47

\bibitem[{Zhu {et~al.}(2015)Zhu, Dong, Stone, \& Rafikov}]{Zhu_2015}
Zhu, Z., Dong, R., Stone, J.~M., \& Rafikov, R.~R. 2015, \apj, 813, 88

\end{thebibliography}

\begin{appendix} 
\section{Forbidden line tracers with ESPRESSO}
\label{sec:espresso}

\begin{figure}  
 \centering
\includegraphics[trim={0 0 0 6cm},clip,width=0.9\hsize]{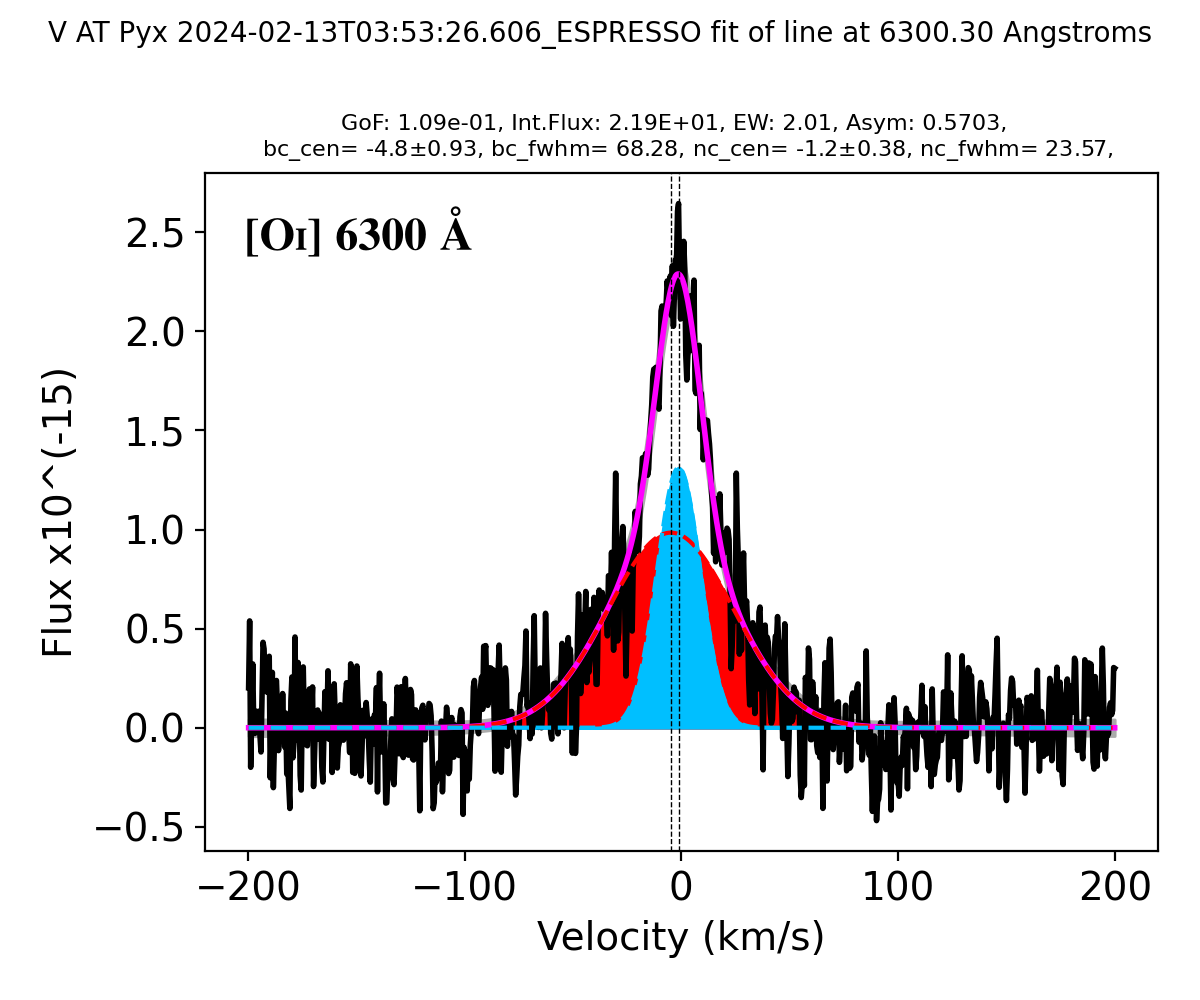}
\includegraphics[trim={0 0 0 6cm},clip,width=0.9\hsize]{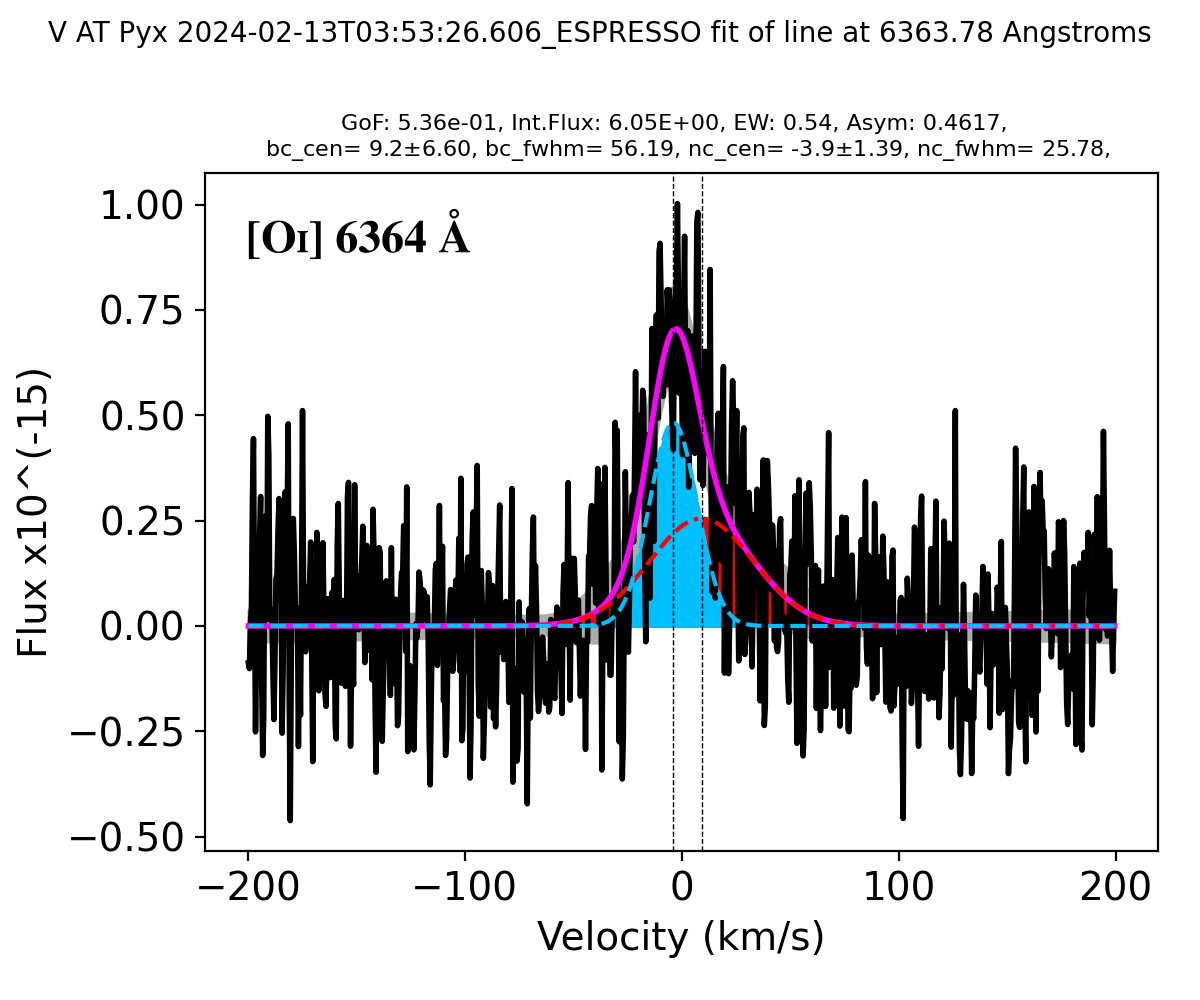}
\includegraphics[trim={0 0 0 6cm},clip,width=0.9\hsize]{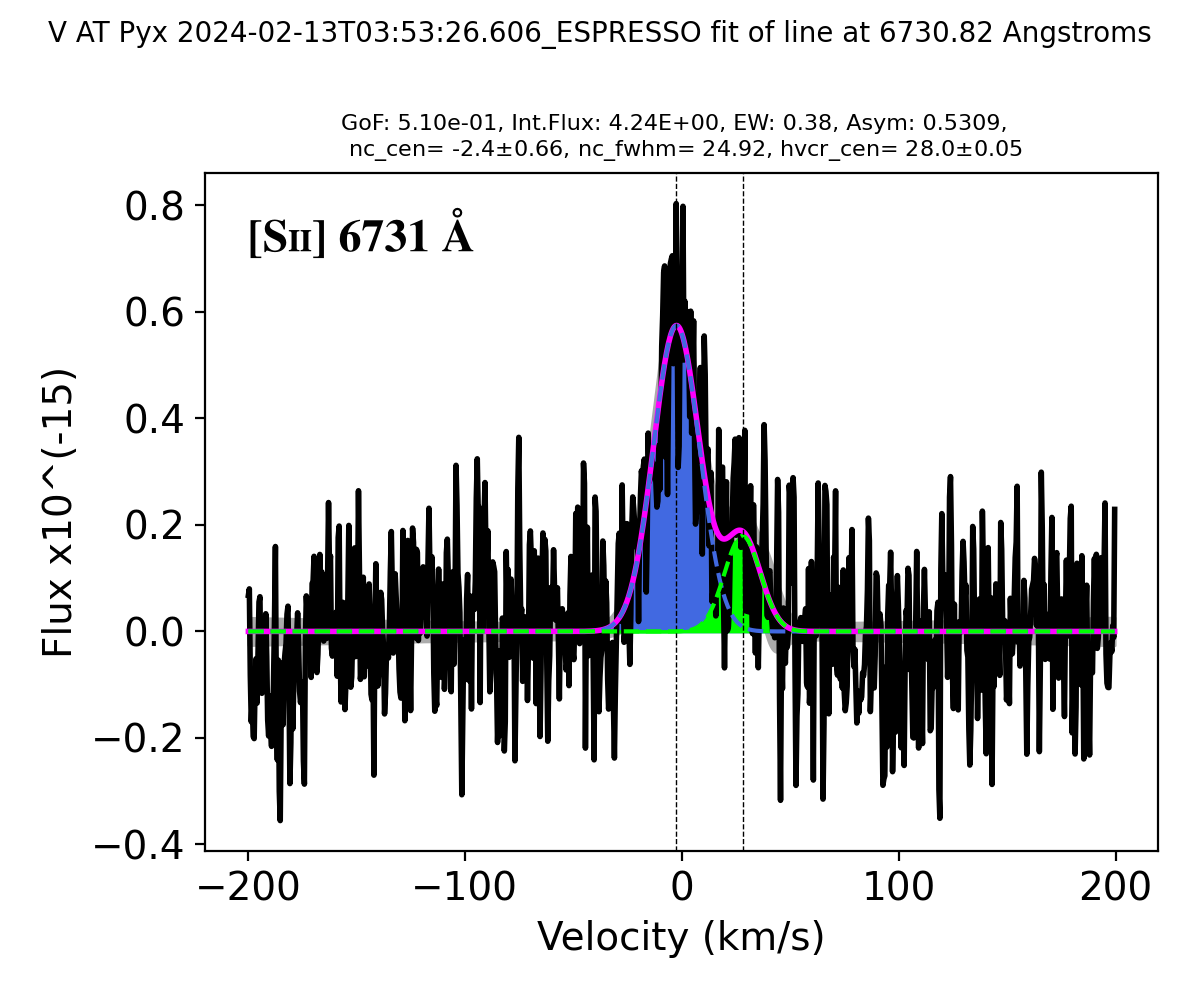}
\caption{Profiles of all forbidden lines detected within ESPRESSO spectra, and fitted using the STAR-MELT package. The colour-coding is as follows: black - observed spectrum; pink - fitted curve to total line profile; light blue - narrow component of LVC; red - broad component of LVC (broad defined as FWHM > 40 km/s); dark blue - LVC with single component; green = HVC (defined as deviating by more than 30 km/s from line centre). \textit{Top:} [O\textsc{i}] 6300 \r{A} fit. \textit{Middle:} [O\textsc{i}] 6364 \r{A} fit. \textit{Bottom:} [S\textsc{ii}] 6731 \r{A} fit.}
\label{fig:espresso}
\end{figure}

ESPRESSO spectra for AT Pyx (ESO programme ID 111.24UB.001) showing the [O\textsc{i}] 6300 \r{A}, [O\textsc{i}] 6364 \r{A}, and [S\textsc{ii}] 6731 \r{A} forbidden lines are presented in figure \ref{fig:espresso}. The spectra were analysed using the STAR-MELT package \citep{Campbellwhite2021}, which included photospheric feature removal, radial velocity calibration and fitting. No other forbidden lines were detected. All three line fits suggest a blueshifted peak characteristic of an internally driven disk wind (see \citealt{Ercolano2010}, \citealt{Weber2020diskwind}, \citealt{Ercolano2022}). Given AT Pyx's location inside a cometary globule, a structure given rise to by external photoevaporation, we expect to find forbidden line tracers for external photoevaporation of the disk. However there are no forbidden line tracers detected across ESPRESSO or XSHOOTER's observing spectral ranges that are strong indicators of external photoevaporation. \cite{Ballabio2023} find that despite [O\textsc{i}] 6300 \r{A} being a forbidden line associated with external photoevaporation, it is extremely difficult to disentangle the external wind contributions to the line flux from the internal wind components due to the external wind components' low velocity. A strong tracer for external photoevaporation, according to \cite{Aru2024} is [C\textsc{i}] 8727 \r{A}, however this falls outside of the spectral range of the ESPRESSO observations, and is not present in the XSHOOTER observations. Hence, our study of forbidden lines from AT Pyx does not detect any clear signatures of external photoevaporation. 

In terms of forbidden lines not detected within our dataset, the 12.81 $\mu$m [Ne\textsc{ii}] line is suggested to be a useful diagnostic for photoevaporation (\citealt{Pascucci_2007}, \citealt{Glassgold2007}, see \citealt{Alexander_2008}). \cite{Gudel2010} find that AT Pyx has a 12.81 $\mu$m [Ne\textsc{ii}] flux of $8.3\pm1.2\times10^{-15}$ erg cm$^{-2}$ s$^{-1}$, which we scale with AT Pyx's GAIA distance of 370$\pm5$ pc to give the star a [Ne\textsc{ii}] 12.81 $\mu$m line luminosity of $1.4\pm0.2 \times 10^{29}$ erg s$^{-1}$ or $3.7\pm0.5\times10^{-5}$ L$_\odot$. \cite{Alexander_2008} suggests that the expected line luminosity from photoevaporative processes on a disk surface are around the $\sim 10^{-6}$ L$_\odot$ mark, and the line flux is also of an appropriate scale to match the predictions of \cite{Glassgold2007}. This provides some evidence for at least the presence of a photoevaporative external disk wind.

\section{A cavity in ALMA continuum emission}
\label{sec:cavity}
\begin{figure*}  
 \sidecaption
\includegraphics[width=6.72cm]{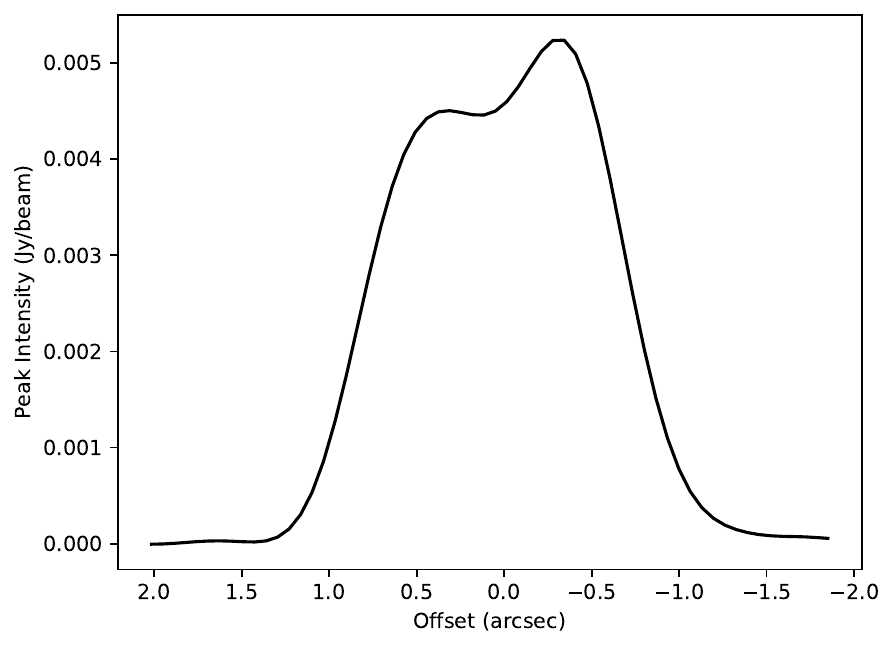}
\includegraphics[width=5.28cm]{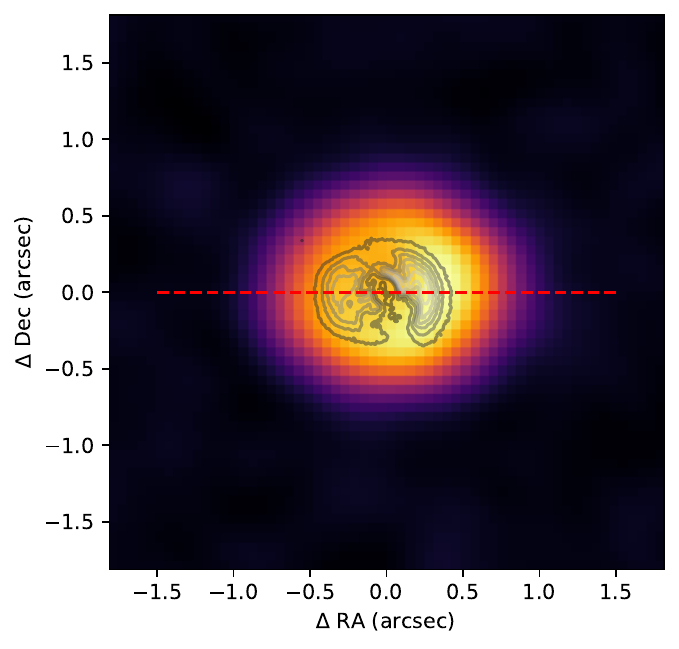}
\caption{\textit{Left:} Radial intensity profile of continuum image. \textit{Right:} Scattered light image contours overplotted on continuum image in greyscale. Dark grey indicates low flux and light grey indicates high flux. The contours serve to show the disk shape in scattered light. The red line indicates the axis of the radial cut over which photometric measurements in the left panel were taken.}
\label{fig:continuumintensityprofile}
\end{figure*}

In figure \ref{fig:continuumintensityprofile} we present the photometric measurements of a radial cut through the continuum emission profile. This shows a dip in brightness interpretable as a cavity in the large dust grain distribution. In the same figure it can be seen that the ansae in scattered light may trace the cavity edges, possibly indicating that the spiral arms bridge a gap between inner and outer disk material.
\section{Motion of spiral arms between SPHERE epochs}
\label{sec:spherespiralmovement}
\begin{figure}  
 \centering
\includegraphics[width=\hsize]{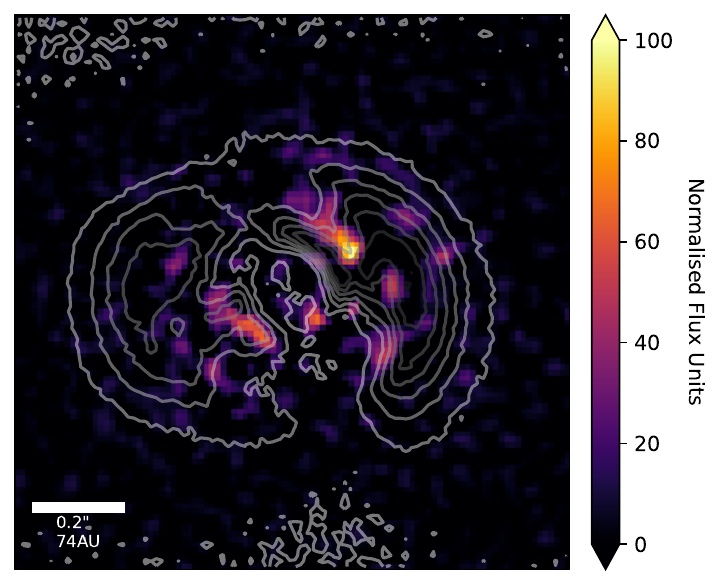}
\caption{Result of subtracting the 2017 SPHERE H-band image of AT Pyx from the 2024 H-band image. Overlaid in greyscale is a contour plot of AT Pyx in H-band to aid with identifying positions of residuals.}
\label{fig:subtraction}
\end{figure}

In order to test whether any of the observed spiral features moved at all between the 2017 and 2024 observations of the system, the 2017 H-band image is subtracted from its 2024 counterpart - with a Gaussian blur applied to both images in order to smooth them - to quickly detect any major changes in the dataset (see figure \ref{fig:subtraction}). In the subtracted image it would appear that the residuals are spotty and trace regions where more flux is detected in the 2024 dataset due to better adaptive optics correction, and there is no notable change in brightness in the position of any of the spirals, indicating they did not move significantly enough in 7 years for a change to be detected within a SPHERE beam ($\sim50$ mas) at this resolution.

\section{Eddy quantities and corner plots}
\label{sec:eddycorner}
{\renewcommand{\arraystretch}{1.2}
\begin{table*}
    \centering
    \caption{Parameters and equations defining Eddy fit}
    \begin{tabular}{|c|c|l|}
    \hline \hline 
    Parameter  & Symbol & Description\\

    \hline
    Centrepoints & $x_0, y_0$ &  Co-ordinates of central star in rotation map in arcseconds\\
    Stellar mass & $M_{star}$ & Mass of star in solar masses\\
    Position angle & $PA$ & Angle of minor axis to y-axis in degrees\\
    Systemic velocity & $v_{LSR}$ & Velocity of whole system with respect to local standard of rest in metres/second\\
    Inclination & $inc$ & Angle about the major axis that the disk plane is rotated in degrees\\
    Aspect ratio & $z_0$ & Gas emission surface height $h_0$ at 1" for the emission surface\\
    Flaring power & $\psi$ & Exponent that determines the power-law for gas surface height\\
    Taper terms & $r_{\text{taper}},q_{\text{taper}}$& Terms describing an exponential taper in emission scale height for the edge of the disk region\\
    
    \hline 
    \multicolumn{2}{|l|}{Velocity} & $v_0 = \sqrt{\frac{GM_{*}r^2}{(r^2+z^2)^{3/2}}}\cos(\phi) \times \sin(inc) + v_{LSR}$ \\

    \multicolumn{2}{|l|}{Emission scale height} & $z(r)=z_0 \times \left( \frac{r}{1''} \right)^\psi \times \exp\left( -\left[\frac{r}{r_{\text{taper}}}\right]^{q_{\text{taper}}}\right)$\\
    \hline \hline
    \end{tabular}
    \newline
    
    \label{tab:eddyparams}
\end{table*}
}
Table \ref{tab:eddyparams} shows all parameters and equations defining eddy fit \citep{Teague_2019}.

\begin{figure*}
   \resizebox{\hsize}{!}
            {
            \includegraphics[width=0.99\hsize]{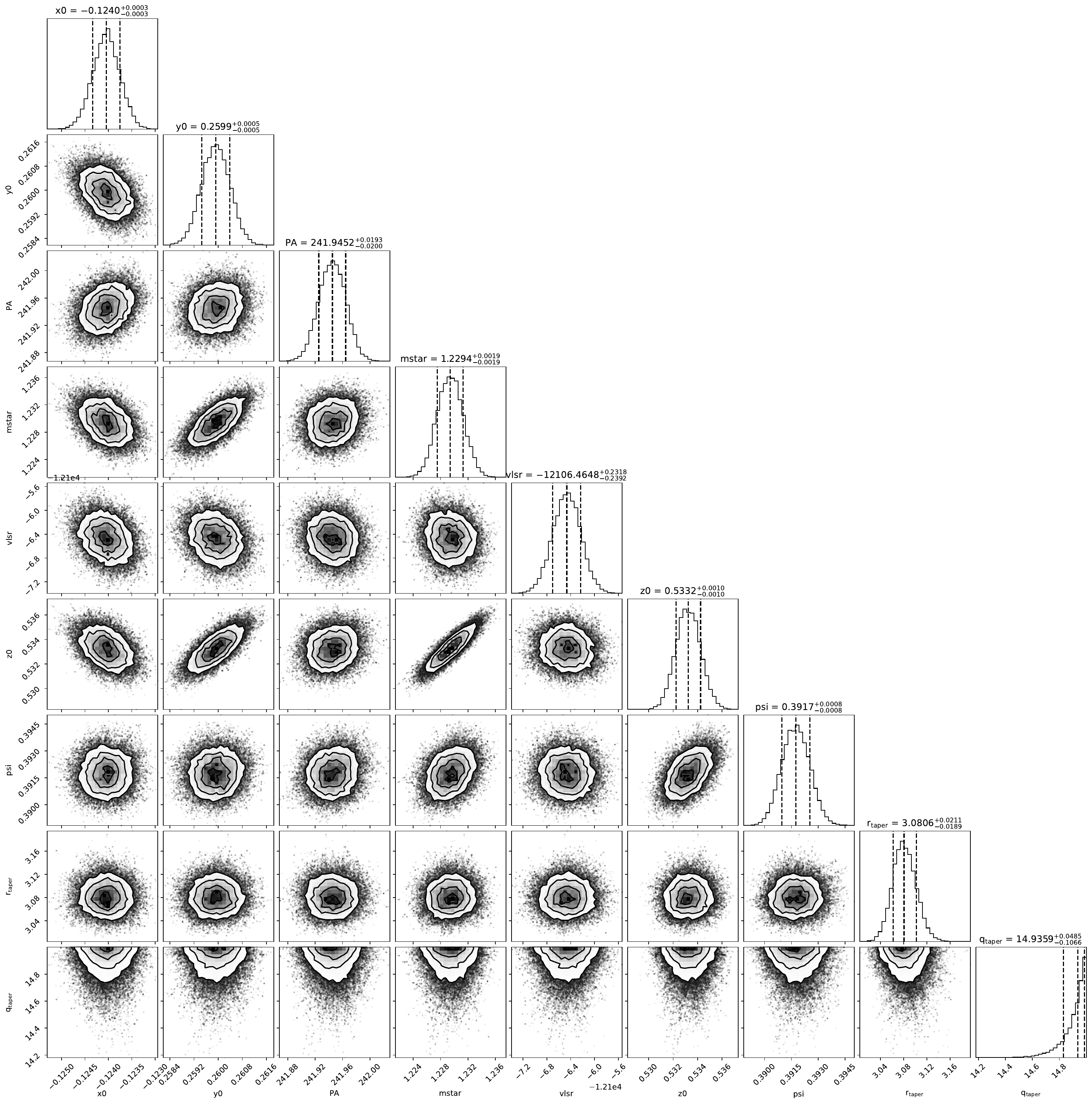}
            }
      \caption{Corner plot showing the final converged values for the Eddy fit. From top to bottom: x-position of disk centre in arcsec, y-position of disk centre in arcsec, position angle in degrees, stellar mass in solar masses, system velocity with respect to the local standard of rest in m/s, pressure scale height at 1'', flaring power, and $r_\text{taper},q_\text{taper}$ (terms describing an exponential taper in scale height at the disk edge).}
\label{fig:eddycorners}
   \end{figure*}

\begin{figure}  
 \centering
\includegraphics[width=\hsize]{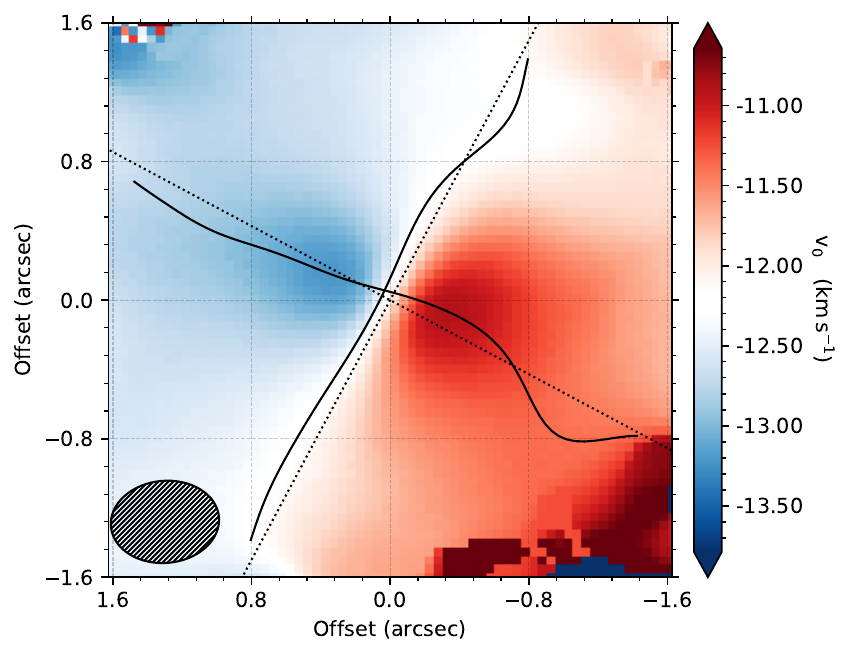}
\caption{Representation of how Eddy measures position angle on a rotation map based on the ALMA gas line data. The solid lines indicates velocity maxima and minima along the axes; the dotted lines indicate the best-fit disk axes.}
\label{fig:positionangle}
\end{figure}

Presented in figure \ref{fig:eddycorners} is a corner plot showing the final converged Eddy disk model parameters after fitting. Each run was subdivided into five iterations of the fitting algorithm in which the converged parameter values at the end of one iteration would be fed in as the inputs to the subsequent iteration so that by the end of the fifth iteration the walkers would have converged on parameter values with minimised error. Each iteration ran with 32 walkers operating over 1500 steps with a 1000 step burn-in phase. Figure \ref{fig:positionangle} shows the disk position angle of 28.06$\pm0.02$$^{\circ}$ as measured by Eddy. Eddy measures position angle as the angle between the y-axis and the major axis as it passes from the origin into the red-shifted region, hence the measured value of $\sim$ 242$^\circ$ is corrected to $\sim$ 28$^\circ$ to refer to the angle between the major axis and the horizontal.

\section{Spiral arms as a result of gravitational instability}
\label{sec:gravitationalinstability}
In terms of another non-planet origin for the spirals observed in the disk, such features can arise as a result of disk gravitational instability (\citealt{Cossins_2009}, \citealt{Dong_2016SpiralInstability}, \citealt{Baehr_2021}). In order to check if AT Pyx's disk is gravitationally unstable, the Toomre Q-parameter \citep{Toomre1964} 
\begin{equation}
    \label{eqn:toomre}
    Q = \frac{c_s\Omega}{\pi G \Sigma}
\end{equation}
is invoked. A disk is held to be sufficiently gravitationally unstable to drive the formation of spiral arms at $Q\le 1.7$ \citep{Durisen2007}. Following the methodology used in \cite{ginski2025diskevolutionstudyimaging}, this parameter can be found by relating gas pressure scale height to sound speed $c_s$: $h_g=c_s/\Omega$, where Keplerian rotational frequency $\Omega = \sqrt{GM_*/R^3}$. 

The pressure scale height of the gas can be found by taking the scale height of the near-infrared scattering surface used for the deprojection as mentioned in section \ref{sec:eddy} and dividing $h_{0,\text{IR}}=16.17$ au by a factor of 4 to compensate for the different heights of the emission surfaces \citep{Chiang2001}. 

Disk surface density $\Sigma=M_{\text{disk}}/2\pi R_{\text{tot}}^2$ relies on the total disk mass obtained in section \ref{sec:almacontinuum} of M$_{\text{disk}}\approx9700$ M$_\oplus$. An even distribution of mass is assumed across the disk surface, which is assigned with a total extent $R_{\text{tot}}=620$ au based on the ALMA \textsuperscript{12}CO emission extent. Thus the newly modified formula for Toomre Q-parameter is now 
\begin{equation}
    \label{eqn:toomre2}
    Q=\frac{2R_{\text{tot}}^2h_0M_*}{M_{\text{disk}} R^3}\left(\frac{R}{R_0}\right)^{\alpha}.
\end{equation}

\begin{figure}  
 \centering
\includegraphics[width=\hsize]{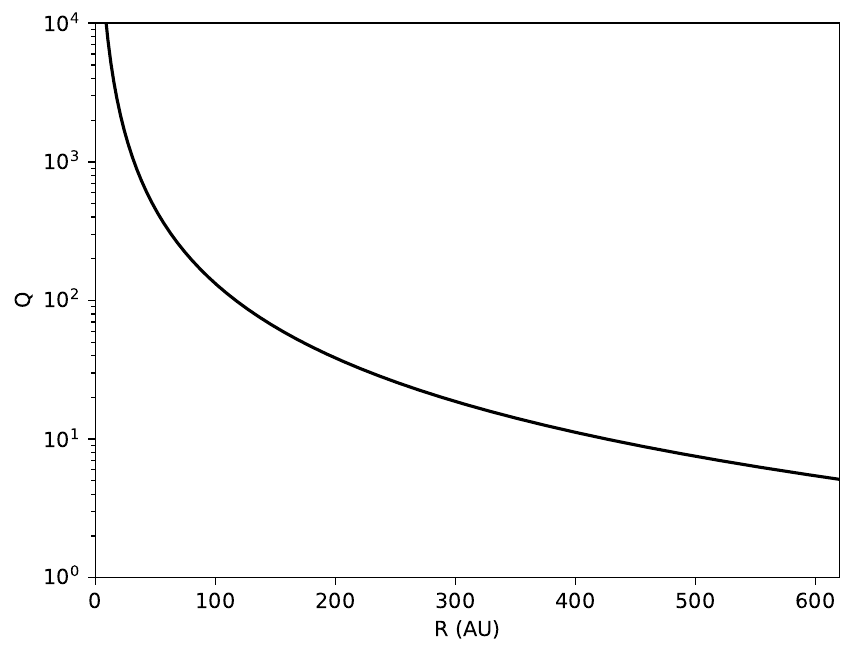}
\caption{Estimated Toomre Q parameter (dimensionless) for gravitational instability as a function of radius.}
\label{fig:toomre}
\end{figure}

As can be seen in figure \ref{fig:toomre}, the appropriate magnitude for gravitational instability is only approached at the furthest edges of the disk, which would seem to indicate that the spiral features of the disk are not a result of gravitational instability.

This approximation is not without flaws; an assumption of a consistent surface density across the disk overestimates the outer disk surface density, however this approximation should still indicate that the inner disk regions can be assumed gravitationally stable given the magnitude of Q for these regions would not drop significantly enough to change the result with a radially dependent surface density model. 

Another potentially significant source of error on this calculation is that the disk dust mass obtained in section \ref{sec:almacontinuum} could be underestimated, as a measurement of dust mass from integrated flux of continuum emission assumes that the disk midplane is completely optically thin. If this is not the case, then the disk dust mass could be underestimated, inflating the Q parameter.

It is found that the disk mass needs to be underestimated by a factor of 1000 for the spiral region of the disk to be within the instability regime. Hence the conclusions here about gravitational instability still hold.

\section{The Gum Nebula moderate-FUV environment and effects on AT Pyx}
\label{sec:fuv}
The primary source of photoevaporative FUV radiation in the cometary globule complex of the Gum Nebula is held to be $\gamma^2$ Vel, a Wolf-Rayet and O star binary \citep{Pozzo2000}. The Vela Supernova Remnant's progenitor Vela XYZ and $\zeta$ Pup, an O star, are also assumed to be contributors to the UV field that gives rise to the comet-like photoevaporative structure of the globules in the local region of the Gum Nebula (\citealt{Kim_2005}, \citealt{Choudhury2009GumNebulaGlobules}).

Estimated FUV field strength at the location of AT Pyx was calculated using similar methodologies to those employed in \cite{valegard2024spherevieworionstarforming}, where the total FUV flux at AT Pyx's position was calculated with 
\begin{equation}
    \label{eqn:fuv}
    F_{\text{FUV,disk}}=\sum_m \frac{L_{\text{FUV,}m}}{4\pi x_{\text{disk}-m}^2}
\end{equation}
with $L_{\text{FUV,}m}$ being the FUV luminosity of each massive star and $x_{\text{disk}-m}$ being the distance between each massive star and AT Pyx. 

$L_{\text{FUV}}$ was found for $\gamma^2$ Vel, $\zeta$ Pup and Vela XYZ by using standard blackbodies and stellar models for O stars from \cite{Pecaut_2013} and integrating over the wavelength range 912-2400 \AA. In order to address the FUV contribution from the no-longer-present Vela supernova remnant progenitor Vela XYZ, similarly to the approach of \cite{Yep2020}, an O star with the same characteristics as $\zeta$ Pup was placed at the location of the Vela Pulsar and put through the same calculations. 

Distances between each star and AT Pyx are calculated trigonometrically with the $x$ and $y$ separations determined simply from on-sky co-ordinates of each source and $z$ separations determined from differences between the Gaia \citep{DISTANCE_gaia_dr3} distance for AT Pyx, the distances for $\zeta$ Pup and $\gamma^2$ Vel from \cite{apellániz2008} and the distance for the Vela Pulsar from \cite{Caraveo2001velapulsardist}. Overwhelmingly the largest source of error on these calculations arise from the error associated with these distances, hence for each star $G_0$ is calculated for the smallest and largest possible separations from AT Pyx $x_{\text{disk-m}}$. 

In current observed configurations the FUV field strength at the location of AT Pyx from equation \ref{eqn:fuv} ranges between 0.93$G_0$ and 28.74$G_0$. When extrapolating the trajectories of $\gamma^2$ Vel and $\zeta$ Pup from the proper motion and radial velocity measurements provided in \cite{Yep2020}, it is also found that $\gamma^2$ Vel is currently experiencing its closest approach to AT Pyx and so its contribution to the FUV field would not have been any greater earlier in AT Pyx's lifetime. $\zeta$ Pup has a trajectory almost parallel to AT Pyx and thus any change in distance is so small that any modelled difference in received FUV flux is within the margin of error and cannot be accurately accounted for. Additionally this extrapolation neglects any changes in luminosity due to stellar evolution.

\begin{figure}  
 \centering
\includegraphics[width=\linewidth]{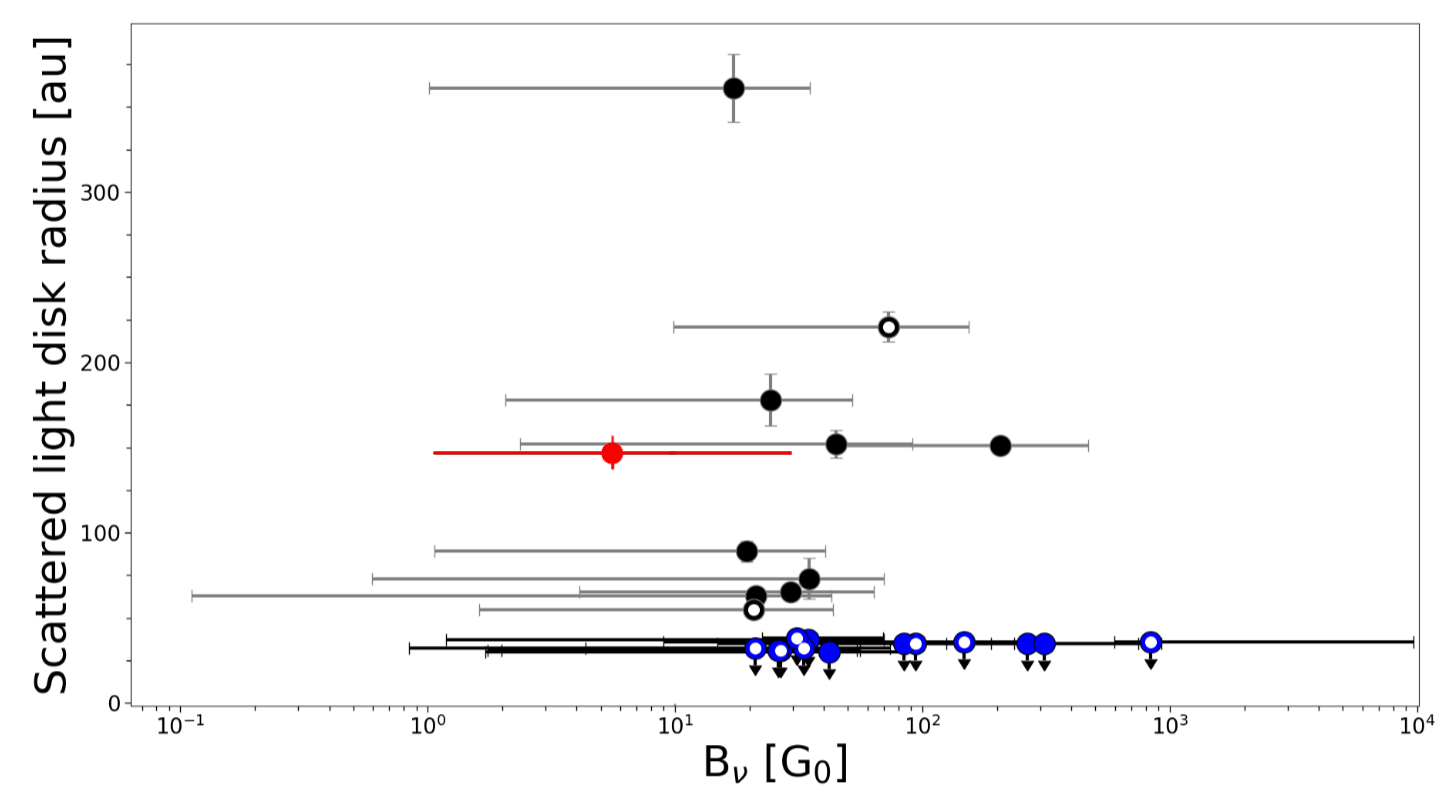}
\caption{Figure 6 in \cite{valegard2024spherevieworionstarforming} showing a sample of stars in the Orion star-forming region with AT Pyx (red) superimposed. Blue indicates no disk detected. No trend is observed.}
\label{fig:valegardfig}
\end{figure}

These results would place AT Pyx at the lower edge of disks grouped by $G_0$, when inserting it among a survey of disk-hosting stars in the Orion star-forming region by \cite{valegard2024spherevieworionstarforming} (see figure \ref{fig:valegardfig}). This survey observes no disks at FUV fields higher than $300G_0$, but aside from this there appears to be no trend between higher FUV field strengths and disk extents. 

Under the assumption that cometary globule structure is a result of radiation driven implosion under a strong FUV field \citep{Bertoldi1989B}, the distinct dense head and wispy tail structure of CG22 - AT Pyx's host globule - would suggest that the FUV field strength at its location is sufficient to induce some level of photoevaporation, but at this point it is difficult to determine what role FUV radiation plays in disk structure for AT Pyx, or even to what extent its position inside the head of CG22 affects its received levels of FUV flux.

\section{Dust masses of AT Pyx and AB Aur}
\label{sec:dustmasses}
\begin{figure}  
 \centering
\includegraphics[width=\hsize]{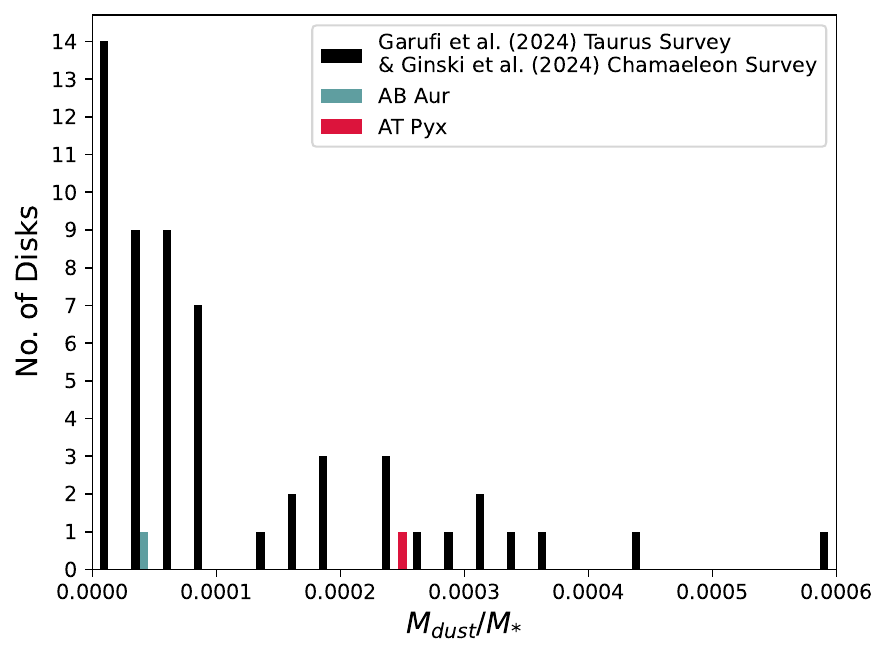}
\includegraphics[width=0.94\hsize]{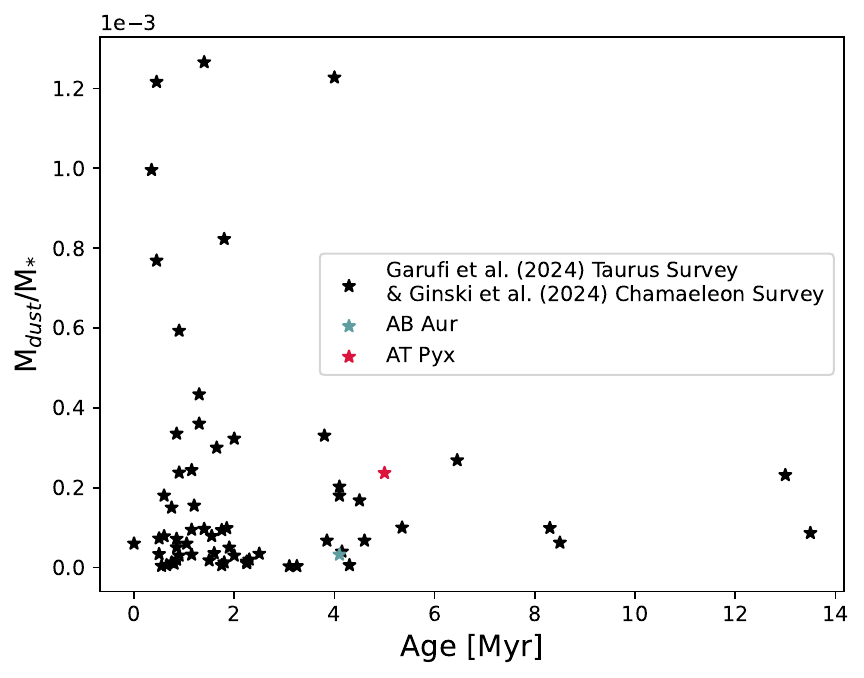}
\caption{Disk dust mass/stellar mass ratios surveyed in Taurus \citep{Garufi2024} and Chamaeleon I \citep{Ginski2024Chamaeleon} star-forming regions, with dust masses for AB Aur \citep{rivièremarichalar2023} and AT Pyx (this paper) inserted.  \textit{Top:} Histogram showing populations for each mass ratio regime. \textit{Bottom:} $M_{\text{dust}}/M_*$ as a function of system age.}
\label{fig:dustmasssurvey}
\end{figure}

Something that could be an indicator of a significant mass contribution from infall is if the dust masses of AB Aur and AT Pyx are greatly inflated in terms of average dust masses of a known disk population. In order to check this, their disk dust/star mass ratios are included among a survey of disks in the Taurus star-forming region \citep{Garufi2024} and a survey of disks in the Chamaeleon I star-forming region \citep{Ginski2024Chamaeleon} in figure \ref{fig:dustmasssurvey}. Both disks are slightly, but not significantly above average in both categories among the 62 stars in the combined surveys.

\end{appendix}

\end{document}